# On the sensitivity to hyperparameters of a Bayesian method to infer selection signatures

Carlos A. Martínez[1*], Ronnal E. Ortiz[2], Nelson A. Cruz[3], Carmen H. Cepeda[4], and Juan Sosa[5].

[1]*Departamento de Producción Animal, Universidad Nacional de Colombia, Sede Bogotá. Carrera 45 # 26-85, Bogotá D.C., Colombia, ORCID:* 0000-0003-2842-5580. [2]*Corporación Colombiana de Investigación Agropecuaria, Sede Central, Km 14 vía Mosquera, Cundinamarca, Colombia, ORCID:* 0000-0003-3977-1462. [3]*Departament Ciències Matemàtiques i Informàtica de la Universitat de les Illes Balears, Cra. de Valldemossa, km 7,5. 07122 Palma (Illes Balears), ORCID:* 0000-0002-7370-5111. [4]*Escuela de Matemáticas y Estadística, Universidad Pedagógica y Tecnológica de Colombia, Facultad Seccional Duitama, Carrera 18 calle 23, Duitama, Boyacá, Colombia, ORCID:* 0000-0002-1317-0776. [5]*Departamento de Estadística, Universidad Nacional de Colombia, Sede Bogotá. Carrera 45 # 26-85, Bogotá D.C., Colombia, ORCID* 0000-0001-7432-4014*:*

*Corresponding author. Email: camartinezn@unal.edu.co
Ciudad Universitaria, Carrera 45 # 26 - 85 Edificio Uriel Gutiérrez
Bogotá D.C.; Bogotá D.C; Postal code: 111321

## Data availability statement.

*Simulated data.*
The simulated data that support the findings of this study are openly available in Zenodo.org at https://doi.org/10.5281/zenodo.21507240, reference number 21507240.
*Real data*.
The real data that support the findings of this study are available from Corporación Colombiana de Investigación Agropecuaria – AGROSAVIA. Restrictions apply to the availability of these data, which were used under license for this study. Data are available from the authors with the permission of AGROSAVIA.

## Conflict of interest disclosure.

There are not financial/commercial conflicts of interests.

## Acknowledgements.

This study was supported by the Colombian government through Ministerio de Agricultura y Desarrollo Rural – MADR and AGROSAVIA. Authors thank Professor Gilberto Pérez from the Mathematics and Statistics School of Universidad Pedagógica y Tecnológica de Colombia – UPTC, Sede Duitama, for useful comments and guidance in proving one of the Theorems.

**Abstract**

Motivated by a real-life problem, namely, inferring genomic regions affected by selection, we studied some properties of a two-step method that performs this task based on Wright's $F_{ST}$. This approach exhibits computational simplicity as an appealing feature. Step one involves fitting a conjugate Bayesian model to find point estimates of $F_{ST}$ and selecting the markers to be considered in step two, where they are grouped using such estimates. Subjective priors are frequently used in this problem and large hyperparameters may be encountered; hence, there is interest in addressing their impact. Preliminary analyses suggested that, the number of markers selected in step one tends to zero when the hyperparameters have the same value and tend to infinity, or are different but tend to infinity and their difference tends to zero. Thus, we studied how hyperparameters affect the posterior mean, variance, and predictive density of allele frequencies, and the posterior distribution of $F_{ST}$, formally. Moreover, an empirical approach based on simulated and real data was performed. Our main findings are summarized in two theorems that set the asymptotic behaviour of the method, providing a foundation to understand the aforementioned phenomena. As to the empirical studies, the groups created in the second step, the estimated posterior means and the number of selected markers showed high sensitivity to hyperparameters. Albeit the mathematical intractability of the posterior distribution of $F_{ST}$, our results revealed asymptotic properties explaining two features of the method that lacked a formal elucidation.

# 1. Introduction

Several approaches have been proposed to infer selection signatures, namely the task of finding genomic regions that are likely to have been influenced by selection. This can be accomplished by studying variation in allele frequencies of these regions across two or more populations. One of the most popular approximations is based on population differentiation inferred via the Wright's fixation index $F_{ST}$ (Lewontin and Karakuer, 1973). Several statistical formulations to estimate this parameter can be found in the literature; for example, frequentist methods are discussed in Weir and Cockerham (1984), Weir and Hill (2002) and Balding (2003), while Bayesian approaches can be found in Holsinger (1999), Beaumont and Balding (2004) and Gianola et al. (2010).

In particular, the later method (hereinafter referred to as the G method) is composed of two steps and features computational simplicity as one of its main advantages. The first step infers the allele frequencies of $L$ biallelic loci across a collection of $R$ subpopulations using a Bayesian model with a product Bernoulli likelihood and, identical and independent Beta priors. There is an $F_{ST}$ for each marker and it is symbolized as $\theta_l, l = 1,2,\ldots,L$. Then, a vanilla Monte Carlo method is used to estimate $F_{ST}$ through the posterior mean. Afterwards, for each marker, a null posterior distribution is found by assuming that allele frequencies in different populations are equal in distribution. Subsequently, the posterior means of $F_{ST}$ computed from the full model (denoted by $\hat{\theta}_l$) are used to detect markers whose $\hat{\theta}_l$ are atypical under the null posterior distribution; those satisfying this condition are considered in the second step where $\{\hat{\theta}_l\}_{l=1}^{L}$ are regarded as "data" and clustered using a finite mixture Gaussian model. These groups are expected to reflect different evolutionary processes occurring in the subpopulations. The aforementioned computational simplicity comes from the use of a simple Monte Carlo integration technique instead of the more complex MCMC algorithms. It is important to point out that MC integration is used due to the mathematical intractability of the posterior distribution of $F_{ST}$.

The G method has been used by several authors studying species of economic relevance (Qanbari et al., 2011; Ma et al., 2014; Ma et al., 2015a; 2015b; Chen et al., 2016; 2018). For example, Ma et al. (2015a) used it to detect selection signatures in pigs, while Chen et al. (2018) went beyond and used it to find candidate genes regulating fat deposition in muscle in a population of Chinese indigenous pigs. Moreover, Qanbari et al. (2011) applied it to detect 236 regions on the bovine genome that were deemed to have experienced recent positive selection. Therefore, the method has proven to be a useful tool in genomic data analysis. Its simplicity makes the G method attractive, but so far, its statistical properties have not been studied, in particular, sensitivity to hyperparameters is a relevant issue in Bayesian analysis (Ghosh et al., 2006).

In the field of animal genetics there is information to use subjective priors using literature reports of allele frequencies or past data from closely related populations, and this approach leads to hyperparameters with large values, as opposed to the case of objective priors like Jeffreys' (Beta(1/2, 1/2)) or Laplace's (Beta(1,1)); furthermore, empirical and hierarchical Bayes approaches may lead to the same scenario. Section 2.2 discusses the need for priors with large hyperparameters in this particular problem. Therefore, there is interest in evaluating the sensitivity of the method to the values of hyperparameters. In that regard, preliminary results from real and simulated data analyses (detailed later) yielded two interesting phenomena that laid the main motivation for this study. The first one occurs when both hyperparameters are equal and this value increases, then the number of selected markers in step 1 decreases towards zero. On the other hand, a marked decrease of the number of selected markers in step 1 is also observed when the hyperparameters are not equal, but their difference tends to zero and their values increase. Therefore, the objective of this study was to assess the impact of hyperparameters on this method through a theoretical approach and complementary empirical studies.

# 2. Methods

## 2.1. Statistical model and parameter estimation

The first step of the method is based on a simple model for biallelic loci. Let $X_{ijlr}$ be a Bernoulli($p_{lr}$) random variable indicating whether the $j^{th}$ copy ($j = 1,2$) of the $l^{th}$ marker ($l = 1,2, \dots, L$) of the $i^{th}$ individual from the $r^{th}$ subpopulation ($r = 1,2, \dots, R;\ i = 1,2, \dots, n_r$) is the reference allele denoted by $A$ (the other marker allele is denoted by $B$). The Bayesian model is as follows:

$$X_{ijlr} | P_{lr} = p_{lr} \overset{ind}{\sim} Bernoulli(p_{lr})$$

$$P_{lr} \overset{iid}{\sim} Beta\left(\frac{1}{2},\frac{1}{2}\right)$$

Hereinafter, this model will be denoted as the full model and the collection of the $R$ subpopulations will be referred to as the population or the complete population. The $Beta\left(\frac{1}{2},\frac{1}{2}\right)$ distribution is a non-subjective prior corresponding to the Jeffreys prior as well as to Bernardo's reference prior (Ghosh et al., 2006). It enjoys certain theoretical properties such as invariance under one-to-one differentiable transformations, first order probability matching and maximizing the Kullback-Leibler divergence between the prior and the posterior distributions (Ghosh et al., 2006). In addition, the Beta family is conjugate for the Bernoulli likelihood; therefore, the posterior is a Beta distribution as well; this is a key feature for the computational simplicity of the G method. Under the null model, it is assumed that $P_{rl} \overset{d}{=} P_l \; \forall\, l = 1,2,\dots,L, \forall\, r = 1,2,\dots,R$, where the symbol $\overset{d}{=}$ represents equality in distribution, this means that, for each marker locus, allele frequencies come from the same population. Here, we allow the model hyperparameters to vary since our target is studying their impact on this inference problem.

Let $a, b \in \mathbb{R}_+$, where $\mathbb{R}_+$ is the set of positive real numbers. Thus, assuming independent and identical $Beta(a, b)$ prior distributions for allelic frequencies, the full model posterior density $\pi(\boldsymbol{P}|\boldsymbol{x})$ is the product of $Beta(n_{Alr} + a, n_{Blr} + b)$ densities, where $n_{Alr}$ and $n_{Blr}$ are the counts of $A$ and $B$ alleles for the $l^{th}$ locus in the $r^{th}$ subpopulation, respectively, $\boldsymbol{P} = (P_{11}, P_{21}, \dots, P_{LR})$, and $\boldsymbol{x}$ the observed data. Therefore, a posteriori, allele frequencies are independent random variables. Let $\boldsymbol{\theta} := (\theta_1, \theta_2, \dots, \theta_L)$; the posterior distribution of $\boldsymbol{\theta}$ is mathematically intractable, which prevents exact inference. Hence, the G Method approximates the posterior mean of $\boldsymbol{\theta}$ using Monte Carlo integration to obtain the point estimator $\widehat{\boldsymbol{\theta}}$. For each $l = 1,2,\dots,L$, it is based on $S$ Monte Carlo samples $\left\{p_{lr}^{(s)}\right\}_{s,r=1}^{S,R}$ and it is computed as the arithmetic mean of $\left\{\theta_l^{(s)}\right\}_{s=1}^{S}$ where

$$\theta_l^{(s)} := \frac{\sum_{r=1}^{R}\left(p_{lr}^{(s)}\right)^2 - \frac{\left(\sum_{r=1}^{R} p_{lr}^{(s)}\right)^2}{R}}{\sum_{r=1}^{R} p_{lr}^{(s)} - \frac{\left(\sum_{r=1}^{R} p_{lr}^{(s)}\right)^2}{R}}.$$

Lastly, under the null model, the posterior density of $\boldsymbol{P_N} := (P_1, P_2, \dots, P_L)$ is the product of $L\; Beta(\sum_{r=1}^{R} n_{Alr} + a\, , \sum_{r=1}^{R} n_{Blr} + b)$ densities.

## 2.2. Why considering large hyperparameter values?

When formulating a Bayesian model for the success probability of a product Bernoulli likelihood, some objective priors are Beta distributions with small values of hyperparameters $a$ and $b$; for instance, as already mentioned, the Jeffreys prior corresponds to a $Beta\left(\frac{1}{2},\frac{1}{2}\right)$, and the Laplace prior to a $Beta(1,1)$ that amounts to a $Uniform(0,1)$. However, the use of subjective priors, or decision-theoretic approximations for prior elicitation, may lead to large values of the hyperparameters.

Moreover, empirical Bayes is a way to deal with uncertainty on the hyperparameters by estimating them using the observed data. In this setting, empirical Bayes leads to inferring the parameters of a Beta-Binomial distribution, which may be large. Furthermore, when inferring allelic frequencies, Martínez et al. (2018) used estimates reported in the literature to find the hyperparameters that permitted to match the prior mean and the prior coefficient of variation, this is an approach for prior elicitation that could be used here and may yield large values of $a$ and $b$; Appendix S1 from Martínez et al. (2018) shows an example of a subjective prior with large hyperparameters. On the other hand, based on decision-theoretic considerations, Martínez et al. (2015) derived Bayes-minimax point estimators of allele frequencies with uniformly smaller variance than the maximum likelihood estimator, and the corresponding hyperparameters could be large since they depend on sample size. On the other hand, hierarchical Bayes is another approach to account for hyperparameters uncertainty. Holsinger (1999) proposed a hierarchical model to infer the $F_{ST}$ using MCMC to approximate the posterior distribution. When examining the algorithm, it is clear that the allele frequencies may sampled from Beta distributions with large hyperparameters.

These concerns pose a question regarding the extent of the influence of hyperparameters on the G method and; therefore, suggest performing a sensitivity analysis (Ghosh et al., 2006) which led to preliminary empirical studies that generated the phenomena described in the introduction, which, as stated above, was the main motivation for this paper.

## 2.3. Empirical sensitivity analysis

Given an inferential problem and a family of prior distributions indexed by a parameter $\boldsymbol{\eta} \in \Omega$ and denoted as $\Pi_{\boldsymbol{\eta}}$, a sensitivity analysis in the Bayesian context looks for determining how much the inferences drawn from the posterior change as $\boldsymbol{\eta}$ varies over a set $\mathcal{F} \subseteq \Omega$ (Ghosh et al, 2006). To this end, two approximations were taken in this study, an empirical one based on a simulation study and the analysis of a real dataset, and a theoretical one that is presented in this section and corresponds to the core of the paper.

In this section we describe the empirical approach to assess the sensitivity to hyperparameters of the G method; to this end, simulated and real datasets were used. Two grids (one for simulated and one for real data) of 54 combinations of $a$ and $b$ were created to perform the analysis, the G method was ran for each one of them and the results were used to assess sensitivity through the following parameters: the number of markers with extreme $F_{ST}$ values (NSM), the number of groups found in step two (NC), adjusted Rand index (ARI) which measures pairwise similarity between groups correcting for chance, and per locus coefficient of variation (CV) of the estimated posterior means of $F_{ST}$ across the grid. Five thousand Monte Carlo samples were used to estimate $F_{ST}$ from the posterior distribution of allele frequencies under the full model and the rule to declare a marker as having an extreme $F_{ST}$ value was as follows. For $l = 1,2,\ldots,L$, declare $\hat{\theta}_l$ as departing from the null model if $\hat{\theta}_l \notin \left[\wp_{\frac{\alpha}{2}}, \wp_{\frac{1-\alpha}{2}}\right]$, where $\wp_{\frac{\alpha}{2}}$ and $\wp_{\frac{1-\alpha}{2}}$ are the $\frac{\alpha}{2}$ and $\frac{1-\alpha}{2}$ quantiles of the null posterior distribution of $\theta_l$, respectively. On the other hand, for the second step, the number of groups was allowed to take values in the set $\{1,2,3,5\}$ and the "optimal" number of groups was chosen using the Akaike Information Criterion as in Gianola et al. (2010).

The need for two grids arose from the fact that the one considered in the simulation could not be used in the real dataset because, in many cases, NSM was zero due to the phenomena stated in the MC Theorem (see the theory section); both grids are provided in supplementary file 4. This analysis involved performing the two steps of the G method over the 54 grid positions and computing the parameters used to assess sensitivity; to this end, an in-house R script (R Core Team, 2021) was created (supplementary file 5). Even though the method requires less computational effort than others, our R code was time consuming (around 7 hours per repetition in the simulated data and 15 hours in the real data described below); therefore, since this study was not concerned with computational efficiency nor scalability, the simulated and real datasets were small (as compared to those found in current statistical-genomics studies).

**Simulated data.**

Genotypes of 1247 individuals distributed in three subpopulations for 1000 biallelic molecular markers were simulated using a forward in time approximation as implemented in the QMSIM software (Sargolzaei and Schenkel, 2013). A historical population comprised of 1000 generations of size 5000 was created and three random samples were taken to generate three subpopulations (S1, S2 and S3) which had different mating systems (assortative for S1 and S2 and random for S3), selection criteria (high breeding values for S1, low breeding values for S2, low phenotypes for S3) and numbers of generations (25 for S1, 20 for S2, and 4 for S3), sample sizes were 506, 441 and 300, respectively; these individuals were chosen from the last generations of each subpopulation. The heritability of the trait was 0.5 and its phenotypic variance was 100, the number of chromosomes was 10, all having a length of 10 centimorgan with 100 markers per chromosome. This procedure was replicated 10 times. The aim of this simulation strategy was creating differences in allele frequencies across subpopulations.

**Real data.**

This dataset was obtained from the bovine genetic improvement programs (BGIP) led by Corporación Colombiana de Investigación Agropecuaria – Agrosavia, Mosquera, Cundinamarca, Colombia. Records came from a sample of 400 bovine from four different groups (100 individuals per group) corresponding to three Colombian creole cattle breeds: Blanco Orejinegro (BON), Romosinuano, and Sanmartinero, and the Brahman breed. It contains a set of 6155 SNP markers from Illumina's BovineLD v1.1 BeadChip that have been used in the BGIP because they are informative to differentiate these breeds and satisfied the following criteria: call rates and minor allele frequencies larger than 0.95 and 0.015, respectively (Burgos, personal communication). Therefore, this set of markers or larger sets containing them have been used in preliminary and ongoing unpublished research focusing on characterizing these genetic groups; for example, Valderrama et al. (2021) studied the BON breed genetic structure. Further edition discarded markers with missing values which led to a dataset with 2368 SNP markers. Considering the scope of this study, we decided to discard markers with missing values as opposed to use imputation because despite of the high accuracy that has been reported for certain methods, we did not want to deal with the uncertainty about imputed genotypes. Part of these records are private and, therefore, this dataset cannot be shared. Details on Agrosavia's genetic improvement program of creole cattle can be found in Burgos and Martínez (2019) and Ramirez et al. (2021).

# 3. Theory

In this section, we develop theory that helps to elucidate the impact of hyperparameters from a mathematical point of view. First, we present two theorems on the limiting behaviour of the posterior distribution of $\boldsymbol{\theta}$ and, then describe some results for functions of the posterior distribution of marker allele frequencies.

### 3.1.1. Limiting behaviour

In this section we present two theorems that provide rigorous explanations of the phenomena related to the impact of $(a, b)$ on the number of markers selected in the first step of the G method. Their proofs rely on some results from real analysis (Rudin, 1976; Bartle, 1991) and probability theory (Resnick, 2003; Shao, 2007) and are presented as a supplementary file.

**Theorem 1 (MC).** Under the G method, if $a = b = \delta$, then, the number of markers with extreme $F_{ST}$ values, tends to zero as $\delta$ tends to infinity.

**Proof.** See supplementary file 1.

**Theorem 2 (MPC).** Let $\Delta \coloneqq a - b$, then, under the G method, the number of markers with extreme $F_{ST}$ values, tends to zero as $\Delta$ tends to zero and $b$ tends to infinity.

**Proof.** See supplementary file 1.

### 3.1.2. Analytical study of two posterior moments and the posterior predictive density of allelic frequencies.

Functions of the posterior distribution of allelic frequencies, namely the posterior means, variances, and predictive densities, are treated as functions of hyperparameters and studied from an analytical perspective. This section focuses on allelic frequencies because of the central role that their posterior distribution plays in the G method and its mathematical tractability. Formally, these are real-valued bivariate functions from $\mathbb{R}_+ \times \mathbb{R}_+$ to proper subsets of $\mathbb{R}_+$. The target is setting some general characteristics of these functions that allow to understand their behaviour which, in turn, brings knowledge about how hyperparameters affect the estimation of allelic frequencies.

*Posterior mean and variance*. In order to study the impact of hyperparameters on the behaviour of the posterior mean and variance of allele frequencies analytically, the gradient vectors as well as the Hessian matrices are derived and some of their properties are proven.

Given the form of the posterior distribution of allele frequencies, $\forall\, l = 1,2,\dots,L,\ \forall\, r = 1,2,\dots,R$ it follows that:

$$\mu_{P_{lr}}^{Post}(a,b) \coloneqq E\left[P_{lr}|\boldsymbol{x}^{o}\right] = \frac{n_{Alr} + a}{2n_r + a + b}\ (\because\ n_{Alr} + n_{Blr} = 2n_r)$$

$$V_{P_{lr}}^{Post}(a,b) \coloneqq Var[P_{lr}|\boldsymbol{x}^{o}] = \frac{(n_{Alr} + a)(n_{Blr} + b)}{(2n_r + a + b)^2(2n_r + a + b + 1)}$$

For $\mu_{P_{lr}}^{Post}$ it can be shown that the gradient vector ($\nabla_{1lr}$) and the Hessian matrix ($H_{1lr}$) are as follows (see supplementary file 2):

$$\nabla_{1lr} = \begin{pmatrix} \dfrac{\partial \mu_{P_{lr}}^{Post}}{\partial a} \\ \dfrac{\partial \mu_{P_{lr}}^{Post}}{\partial b} \end{pmatrix} = \begin{pmatrix} \dfrac{n_{Blr} + b}{(2n_r + a + b)^2} \\ -\dfrac{n_{Alr} + a}{(2n_r + a + b)^2} \end{pmatrix}$$

$$H_{1lr} = \frac{1}{(2n_r + a + b)^3}\begin{bmatrix} -2(n_{Blr} + b) & n_{Alr} - n_{Blr} + a - b \\ n_{Alr} - n_{Blr} + a - b & 2(n_{Alr} + a) \end{bmatrix}$$

Notice that $\frac{\partial \mu_{P_{lr}}^{Post}}{\partial a} > 0\ \forall\, a, b\ \in \mathbb{R}_+$ and $\frac{\partial \mu_{P_{lr}}^{Post}}{\partial b} < 0\ \forall\, a, b\ \in \mathbb{R}_+$, that is, for $b$ arbitrary but fixed, $\mu_{P_{lr}}^{Post}$ is strictly increasing in $a$, and for $a$ arbitrary but fixed, it is strictly decreasing in $b$. Now,

$$\nabla_{1lr} = \mathbf{0}_{2\times 1}$$

$$\Rightarrow \frac{1}{(2n_r + a + b)^2}\begin{pmatrix} n_{Blr} + b \\ -(n_{Alr} + a) \end{pmatrix} = \begin{pmatrix} 0 \\ 0 \end{pmatrix}$$

$$\Rightarrow \begin{pmatrix} a \\ b \end{pmatrix} = -\begin{pmatrix} n_{Alr} \\ n_{Blr} \end{pmatrix}$$

which leads to a contradiction because $n_{Alr}, n_{Blr} \in \mathbb{Z}_*$, (the set of non-negative integers) and $a, b \in \mathbb{R}_+$. Consequently, the function has no stationary points. Furthermore, the Hessian matrix is indefinite. To see why is that, notice that the determinant of $H_{1lr}$ is

$$|H_{1lr}| = \frac{-4(n_{Blr} + b)(n_{Alr} + a) - (n_{Alr} - n_{Blr} + a - b)^2}{(2n_r + a + b)^6}$$

hence $|H_{1lr}| < 0 \; \forall \; a, b > 0$, furthermore, since $H_1$ is a real symetric matrix its eigenvalues are real; moreover, their product is equal to $|H_1|$ (Harville, 1997), consequently, since $H_{1lr}$ has dimension $2 \times 2$, its eigenvalues have different signs and consequently, this matrix is indefinite, so this criterion does not permit to define if the function is covex or concave. The case of $V_{P_{lr}}^{Post}$ is more complex and it is summarized in propositions 1 and 2

**Proposition 1**. The function $V_{P_{lr}}^{Post}$ has no stationary points. Also, for $a, b > 0, n_{Alr}, n_{Alr} \in \mathbb{Z}_*$, let us define the following sets:

$$\mathcal{D}_{lr1} \coloneqq \{(a, b, n_{Alr}, n_{Blr}) \colon n_{Alr} + a > n_{Blr} + b\}$$

$$\mathcal{D}_{lr2} \coloneqq \{(a, b, n_{Alr}, n_{Blr}) \colon n_{Blr} + b > n_{Alr} + a\}$$

$$\mathcal{I}_{lr1} \coloneqq \{(a, b, n_{Alr}, n_{Blr}) \colon n_{Blr} + b \geq 2(n_{Alr} + a)\}$$

$$\mathcal{I}_{lr2} \coloneqq \{(a, b, n_{Alr}, n_{Blr}) \colon n_{Alr} + a \geq 2(n_{Blr} + b)\}$$

Then, $\forall \; l = 1,2,..,L; \forall \; r = 1,2,\dots,R$, for a genetic structure of the $r^{th}$ subpopulation for the $l^{th}$ marker such that $(a, b, n_{Alr}, n_{Blr}) \in \mathcal{D}_{lr1}, V_{P_{lr}}^{Post}$ is strictly decreasing in $a$ for fixed $b$ in this subset of its domain. Moreover, the posterior variance is strictly increasing in $a$ for fixed $b$ in the subset of its domain that along with the genetic structure of the $r^{th}$ subpopulation for the $l^{th}$ marker satisfy $(a, b, n_{Alr}, n_{Blr}) \in \mathcal{I}_{lr1}$, which is a necessary and sufficient condition. Similarly, $\forall \; l = 1,2,..,L; \forall \; r = 1,2,\dots,R$, if the genetic structure of the $r^{th}$ subpopulation for the $l^{th}$ marker is such that $(a, b, n_{Alr}, n_{Blr}) \in \mathcal{D}_{lr2}$, then $V_{P_{lr}}^{Post}$ is strictly decreasing in $b$ for fixed $a$. Finally, $V_{P_{lr}}^{Post}$ is strictly increasing in $b$ for fixed $a$ in the subset of its domain that along with the genetic structure of the $r^{th}$ subpopulation for the $l^{th}$ marker are such that $(a, b, n_{Alr}, n_{Blr}) \in \mathcal{I}_{lr2}$, which is a necessary and sufficient condition.

**Proof**: See supplementary file 2

**Proposition 2**. Let $H_{2lr}$ be the Hessian matrix of $V_{P_{lr}}^{Post}$ and $K_{x,a,b} \coloneqq \frac{2n_r+a+b}{2n_r+a+b+1} + 1$, if $|H_{2lr}| > 0$ then the posterior variance function is strictly convex in the subsets of its domain where $\frac{n_{Alr}+a}{n_{Blr}+b} \in \left(K_{x,a,b}^{-1}, K_{x,a,b}\right)$ or $\frac{n_{Blr}+b}{n_{Alr}+a} \in \left(K_{x,a,b}^{-1}, K_{x,a,b}\right)$. Also, the Hessian matrix is indefinite in the mutually exclusive sets where $(a, b)$ and the population structure satisfy $\frac{n_{Blr}+b}{n_{Alr}+a} > K_{x,a,b}$ or $\frac{n_{Alr}+a}{n_{Blr}+b} > K_{x,a,b}$.

**Proof**: See supplementary file 2.

*Posterior predictive density.* The posterior predictive density is used in Bayesian model selection and checking (Ghosh et al., 2006). Thus, deriving this density and studying it as a function of $a$ and $b$ is useful to analyse the impact of hyperparameters. In general, for an absolutely continuous prior distribution $\Pi$ with density $\pi$, a parameter $\eta$ with parametric space $\Omega$, observed data $\boldsymbol{x}^{\sigma}$ and "future" data $\boldsymbol{x}$; under the assumption of conditional independence of $\boldsymbol{x}$ and $\boldsymbol{x}^{\sigma}$ given $\eta$, the posterior predictive density is defined as:

$$m_{\pi}(\boldsymbol{x}|\boldsymbol{x}^{\sigma}) = \int_{\Omega} f\,(\boldsymbol{x}|\eta)\pi(\eta|\boldsymbol{x}^{\sigma})d\eta$$

In this case: $\Omega = (0,1) \times (0,1) \times \cdots \times (0,1)$ and $\eta = \{P_{lr}\}_{l,r\,=\,1}^{L,R}$ thus,

$$m_{\pi}(\boldsymbol{x}|\boldsymbol{x}^{\sigma}) = \int_{\Omega} \prod_{r=1}^{R} \prod_{l=1}^{L} p_{lr}^{n_{Alr}}(1-p_{lr})^{n_{Blr}} \times \frac{p_{lr}^{n_{Alr}^{\sigma}+a-1}(1-p_{lr})^{n_{Blr}^{\sigma}+b-1}}{B(n_{Alr}^{\sigma}+a, n_{Blr}^{\sigma}+b)} d\boldsymbol{p}$$

where $B(\cdot,\cdot)$ is the Beta function. Let $NC = \prod_{r=1}^{R}\prod_{l=1}^{L}\frac{1}{B(n_{Alr}^{\sigma}+a,n_{Blr}^{\sigma}+b)}$, then

$$m_{\pi}(\boldsymbol{x}|\boldsymbol{x}^{\sigma}) = NC \prod_{r=1}^{R} \prod_{l=1}^{L} \int_{0}^{1} p_{lr}^{n_{Alr}+n_{Alr}^{\sigma}+a-1}(1-p_{lr})^{n_{Blr}+n_{Blr}^{\sigma}+b-1}\, dp_{lr}.$$

Now, observing that we are integrating the kernels of Beta densities and using the recurrence property of the Gamma function $\Gamma(x+1) = x\Gamma(x), x > 0$, it can be shown (see supplementary file 3) that

$$m_{\pi}(\boldsymbol{x}|\boldsymbol{x}^{\sigma}) = \prod_{r=1}^{R} \prod_{l=1}^{L} \frac{\prod_{i=1}^{n_{Alr}}(n_{Alr}+n_{Alr}^{\sigma}+a-i)\prod_{i=1}^{n_{Blr}}(n_{Blr}+n_{Blr}^{\sigma}+b-i)}{\prod_{i=1}^{2n_r}(2(n_r+n_r^{\sigma})+a+b-i)}.$$

It is worth mentioning that the dependency on hyperparameters is implicitly expressed through the subindex $\pi$ which represents the prior density that, in this case, is indexed by $(a, b)$. Some analytical results on the behaviour of the posterior predictive density are provided in supplementary file 3.

# 4. Results

## 4.1. Simulation study

For some loci, the posterior means of $\theta_l$ showed considerable sensitivity to hyperparameters since their across-grid CV varied from 0.5 to 1.5; out of the 1000 simulated markers, 63 had an across-grid CV greater or equal than 1 and 673 had an across-grid CV greater or equal than 0.7. Figure 1a shows the empirical distribution of the across-grid CV of the posterior means for the 1000 simulated markers throughout the ten replicates. Furthermore, NSM exhibited high variation across the 54 hyperparameter values; across-replicates mean varied from 111.3 to 944.5. Figure 2 presents its behaviour along the grid of hyperparameter values, the boxplot built from the ten replicates is portrayed at each position.

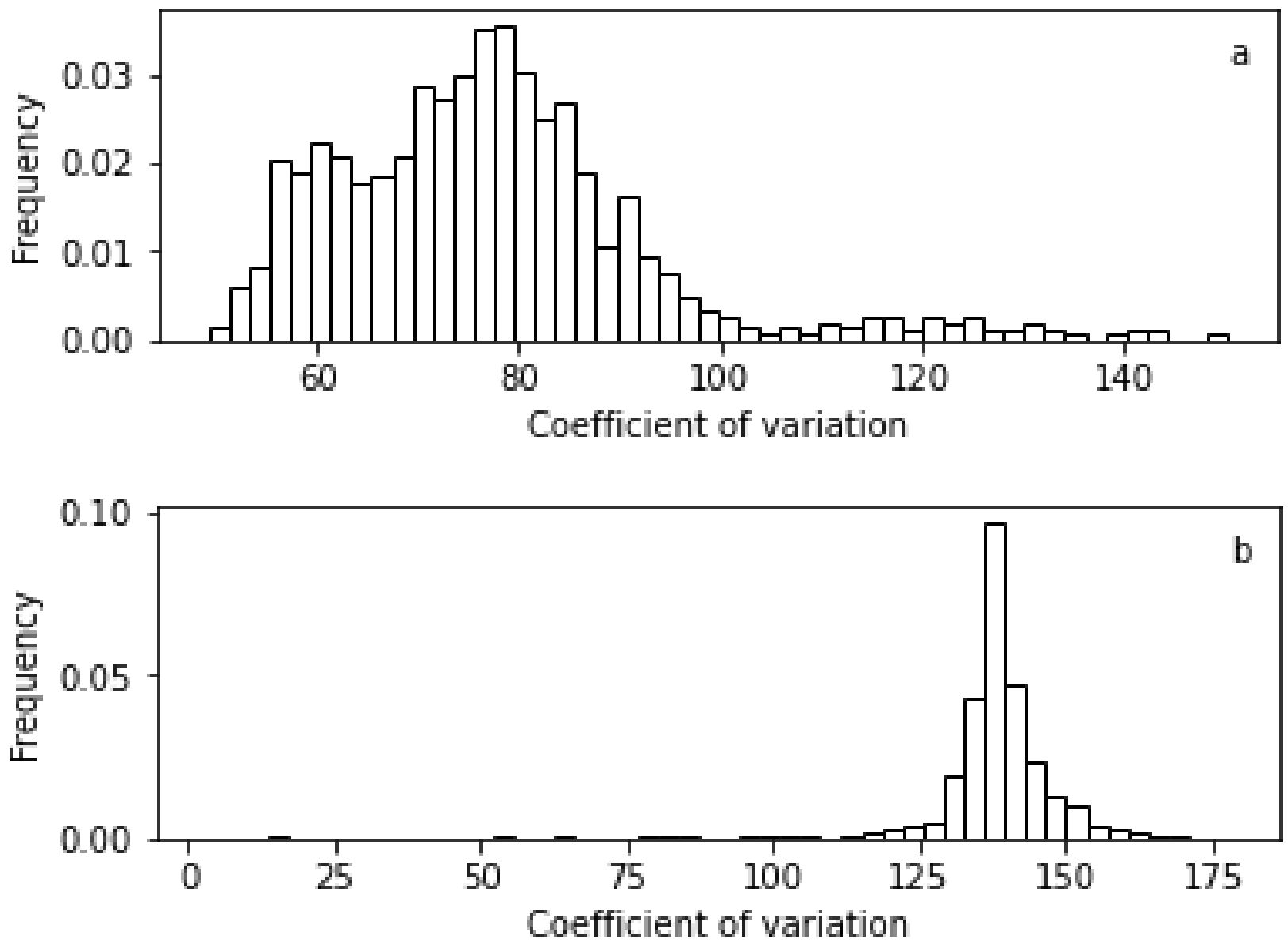

**Figure 1.** Empirical distribution of the coefficient of variation of posterior means of $F_{ST}$. **a**. Across-replicates coefficient of variation of the posterior means of $F_{ST}$ for the 1000 simulated marker genotypes. **b**. Coefficient of variation of the posterior means of $F_{ST}$ for the 2368 real marker genotypes.

On the other hand, NC displayed low sensitivity, the most frequent value was 5. The overall mean across hyperparameter values and replicates was 4.74 with a CV of 0.09. It is worth mentioning that NC 1 was never selected, and 2 was chosen only 4 times out of 540 analysis (54 hyperparameter values and 10 replicates).

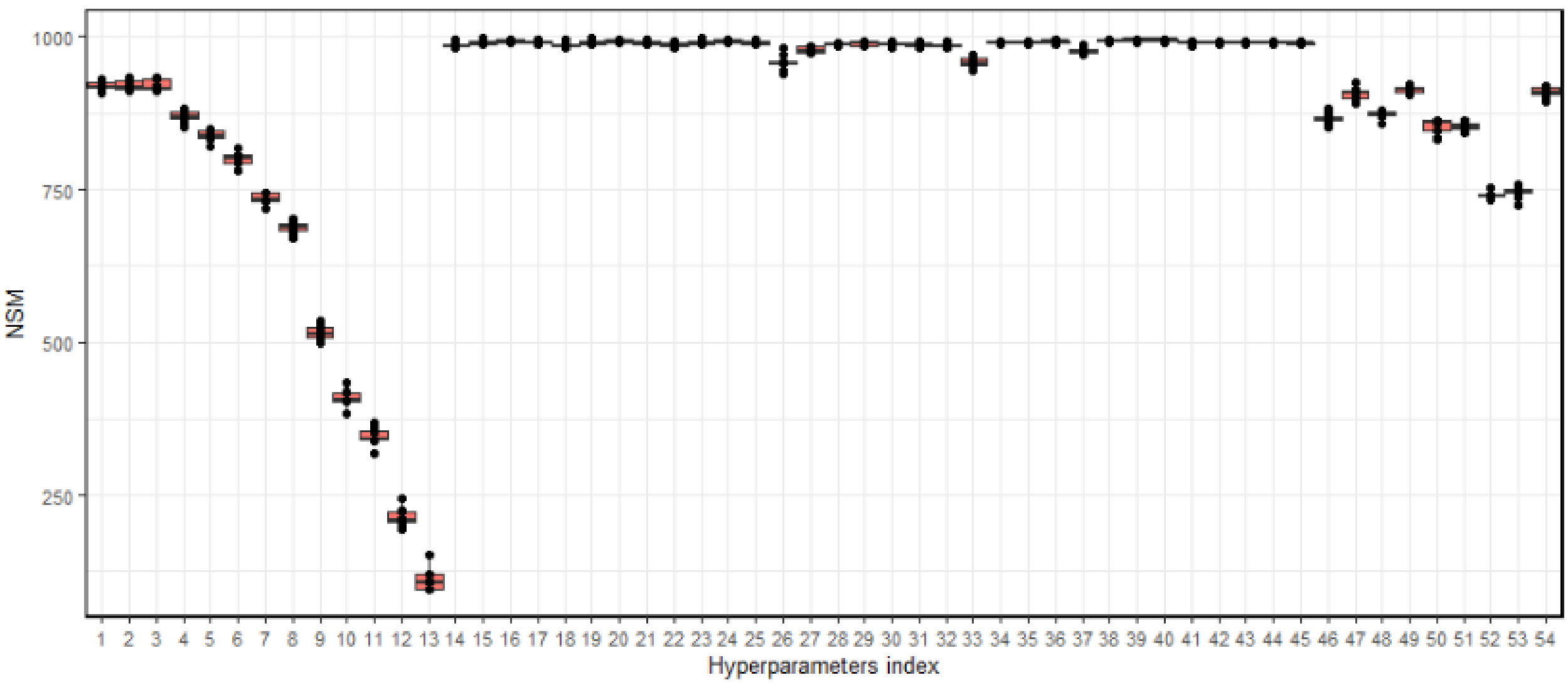

**Figure 2.** Behaviour of the number of selected markers (NSM) across the grid of hyperparameter values in the simulated data. The boxplot built from the ten replicates is portrayed at each position.

Additionally, ARI varied from 0.05 to 0.98, which suggests sensitivity to hyperparameters, also, it means that pairwise similarity between groups of markers ranged from what would be expected by chance (values close to cero) to high similarity (values close to 1). Figure 3a shows the histogram of ARI for the $\binom{54}{2}$ pairs of groupings across all ten replicates.

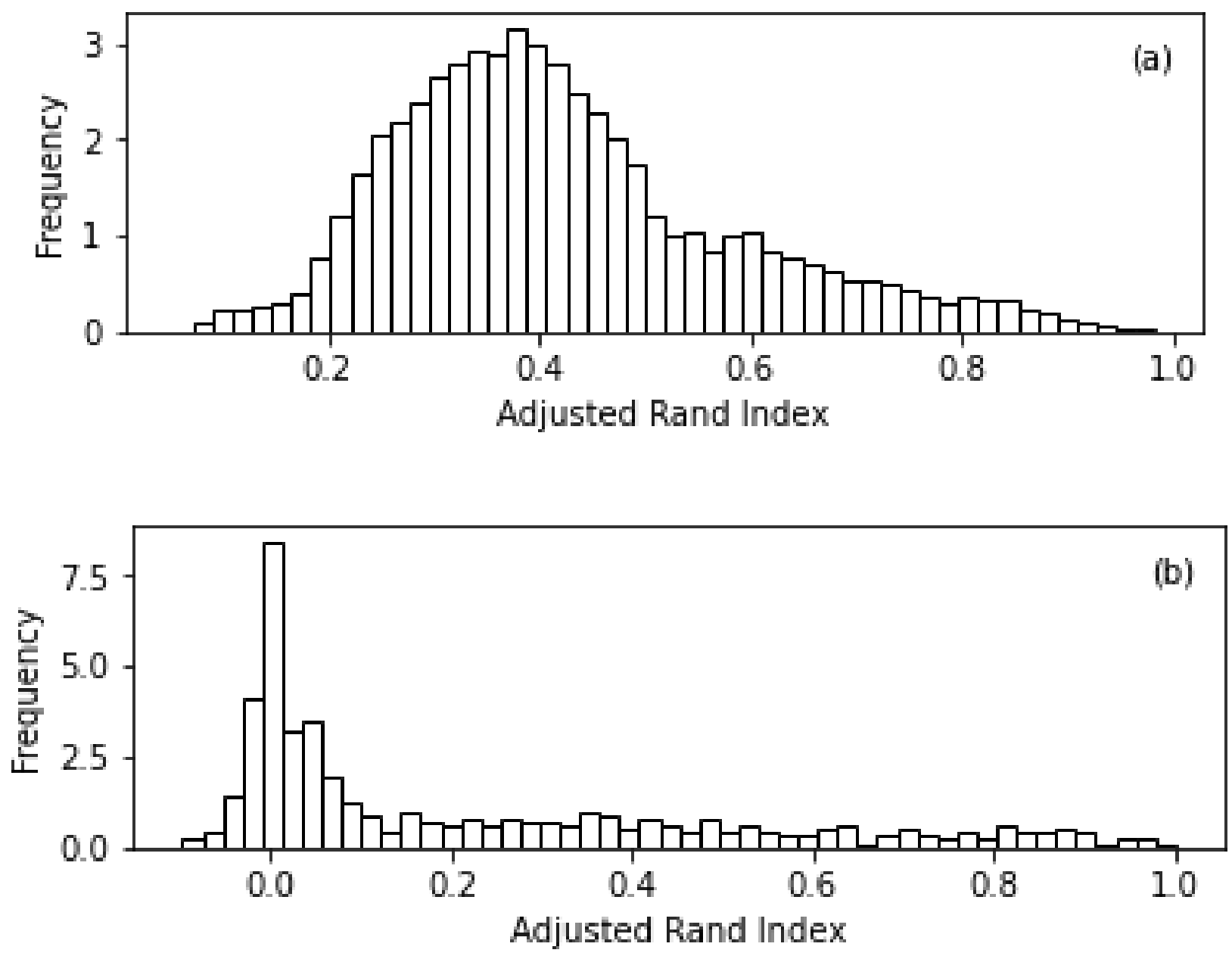

**Figure 3** Histogram of the adjusted Rand index for all possible pairs of groupings. **a.** Simulated data. The histogram covers results from all ten replicates. **b.** Real data.

### 4.2. Real data

Overall, the results from the bovine SNP data were similar to those found in the simulation study with the exception of NC. Figure 4 shows the behaviour of NSM along the grid of hyperparameter values. Notice that, in this case, NSM got values as low as 3 when $a = b = 7500$, position 13 in the grid.

Furthermore, the posterior mean of $F_{ST}$ also suggested high sensitivity to the choice of hyperparameters for most of the loci. The CV computed across the 54 grid values varied from 0.03 to 1.78, a broader range than in the simulation study; the empirical distribution of this CV for all markers is presented in Figure 1b. Also, ARI had a broader range than in the simulation study, it took values from -0.10 to 1; therefore, in this case pairwise similarity between groups varied from slightly less than expected by chance alone to complete similarity (Figure 3b).

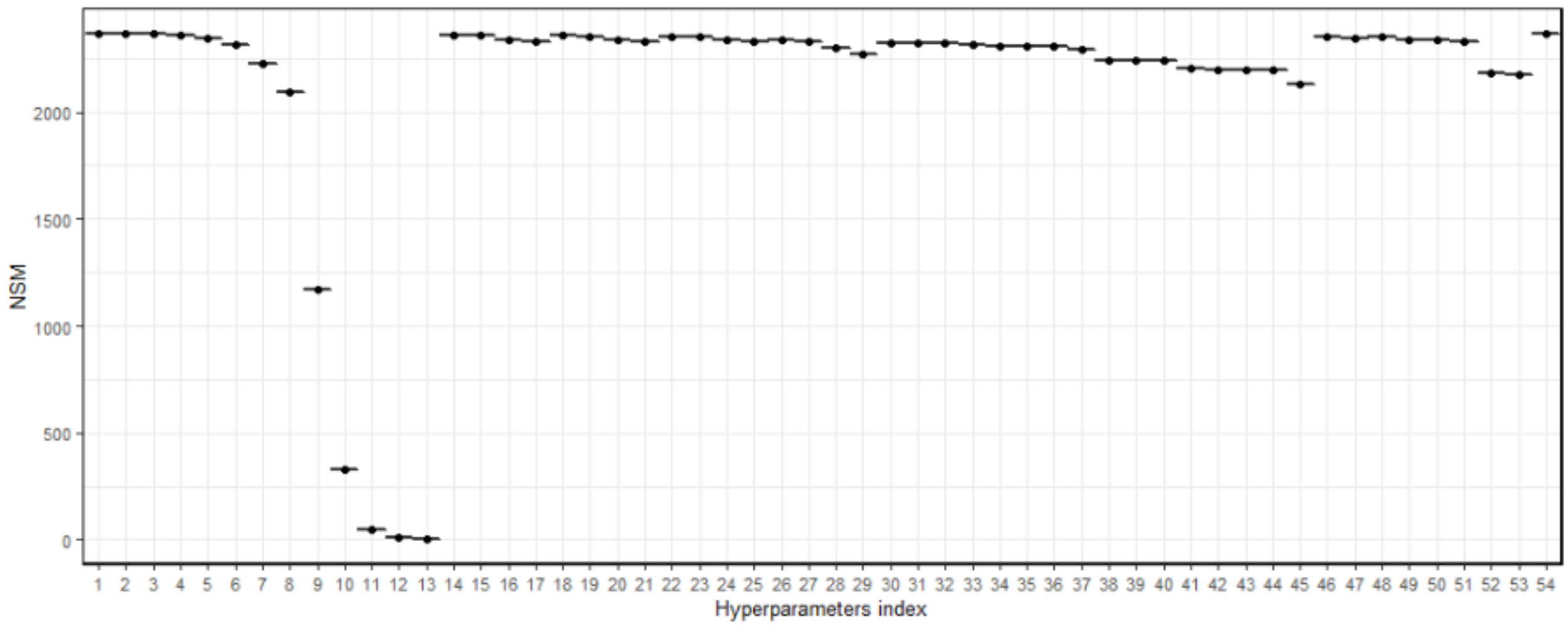

**Figure 4.** Behaviour of the number of selected markers (NSM) across the grid of hyperparameter values in the real dataset.
When the hyperparameters had the same value, NSM decreased as this value increased, recall that this was the phenomenon that motivated this study.

On the other hand, NC exhibited higher variation across the grid than it did in the simulation, the most frequent value was 1 (selected 63% of times); the CV of NC was 0.74. Thus, as opposed to the results from simulated data where one group was never chosen, in the cattle dataset it was the most frequent result meaning that all $F_{ST}$ values were found to be similar enough as to be classified in the same group. Additionally, this explains why ARI reached the value of 1.

# 5. Discussion

## 5.1. Theoretical issues

The central role that allele frequencies play when inferring selection signatures using $F_{ST}$ and the mathematical tractability of their posterior distribution motivated us to propose it as a target for an analytical study of sensitivity to hyperparameters. The results for the posterior mean function and propositions 1 to 3 provide some knowledge on how hyperparameters affect allele frequencies estimation. For example, the posterior mean function has no global maximum nor minimum points, therefore, the function has to be evaluated at the boundaries. In view of the form of the gradient vector, it follows that the larger $a$ and the smaller $b$, the larger the posterior mean. Now, since $0 < a, b < \infty$, the parameter space is unbounded from above and has 0 as infimum. The pragmatic implication of this result is that as $b$ approaches zero and $a$ increases, the posterior mean gets closer to its supremum (1); conversely, as $a$ approaches zero and $b$ increases, the posterior mean approaches its infimum (0). This behaviour is preserved across loci, but the functional form of $\nabla_{1lr}$ shows that the rates of change (partial derivatives) depend on the population structure for each marker, in particular, these can be shown to depend on the observed allele frequency.

As to the posterior variance function $V_{P_{lr}}^{Post}$, it displayed a more complex behaviour as can be noticed in propositions 1 and 2. This function describing dispersion around the posterior mean at a given marker and subpopulation has no stationary points, besides, it is worthwhile noticing that the sets $\mathcal{D}_{lr1}$ and $\mathcal{D}_{lr2}$ are disjoint as well as $\mathcal{I}_{lr1}$ and $\mathcal{I}_{lr1}$ which implies that $V_{P_{lr}}^{Post}$ cannot be simultaneously increasing/decreasing in $a$ and $b$. Regarding the posterior predictive density, using properties of the Gamma function (Artin, 1964; Kosmala, 2004) and the fact that $n_{Alr}, n_{Blr}$ and $2n_r$ are positive integers, the posterior predictive density can be expressed as products and ratios not involving the Gamma function, thus getting rid of the need for evaluating it (supplementary file 3). Its behaviour is conditioned by the relationship between hyperparameters and the observed allele configuration as stated in Proposition 3. However, this relationship is more complex than it was for the posterior mean and variance functions. Proposition 3 mentions necessary conditions for $S_{lrb} < 0$ or $S_{lra} < 0$, namely, allele $B$ of the $l^{th}$ marker not being fixed in the sample from the $r^{th}$ subpopulation, and allele $A$ of the $l^{th}$ marker not being fixed in the sample from the $r^{th}$ subpopulation, respectively. This is the typical situation in genomic data analysis because records are edited in such a way that loci with fixed alleles are excluded from the analysis.

Theorems MC and MPC reveal novel limiting properties of the posterior distribution of the fixation index under the G method. The former shows that when the hyperparameters have the same value and it grows, the number of markers selected for step two decreases and eventually is zero. Clearly, this is an asymptotic property, but it does not follow the typical setting studied in theoretical statistics where it is the sample size $n$ that tends to infinity as opposed to a parameter of the distribution. On the other hand, the MPC theorem presents a quite similar limiting performance in a more general but complex context, that is, the hyperparameters have different values (typical situation in a subjective Bayes analysis) and simultaneously, their values grow, and their difference tends to zero. Hence, the usual techniques to show convergence of sequences of random variables could not be applied, so several results from real analysis and probability were combined to prove this result (supplementary file 1). The key feature in both theorems is proving convergence of $F_{ST}$ to a constant under both the full model, and the null model. Moreover, for the special case $a = b = \delta \in \mathbb{Z}_+$, we proved almost sure convergence of $F_{ST}$, while in the general case $(a, b) \in \mathbb{R}_+^2$ we established convergence in probability, a weaker form of convergence (Resnick, 2003).

Another relevant feature of these theorems is the following. When selecting markers for the second step of the G method, Gianola et al. (2010) considered extreme values of $F_{ST}$ in both tails of the null distribution because balancing selection may cause similarity of allele frequencies across subpopulations leading to small values of the fixation index (Cavalli-Sforza, 1966; Beamunt and Balding, 2004), a fact that has been outlined by other authors studying selection signatures (López et al., 2015) and population differentiation (Brandt et al., 2018). However, some studies consider divergent selection which only induces $F_{ST}$ values close to 1; this means that the rule to detect extreme $F_{ST}$ values changes to a single (right) tailed one. In this case, these theorems are still valid since convergence to 1/2 under the full and null models still holds; this is an appealing feature.

In summary, for arbitrary but fixed $n$, as $a$ and $b$ get close and increase, the number of markers selected in step one, and consequently, the number of markers declared as influenced by selection gets smaller and eventually is zero. In other words, $a$ and $b$ can be chosen in such a way that no markers are found to be influenced by selection. This is an artifact that results from an undesirable mathematical property of the model used in step 1 of the G method.

Some topics not covered in this study that can be addressed in future work are: 1) the study of the limiting behaviour of the method when the hyperparameters and the sample size grow, a much more challenging problem from the mathematical point of view, 2) the impact of the relationship between population structure and hyperparameters on the rates of convergence of $\boldsymbol{\theta}$ under theorems MC and MPC.

### 5.2. Data analysis

These analyses were complementary to the core approach of this paper, that is, the theoretical treatment of the problem. Similar results were found in two datasets with a different number of subpopulations, individuals, markers and more importantly, a very different origin as one was real and the other one was simulated. Among such evidence we can find the variability of the estimated posterior mean of the fixation index across grids as suggested by its CV (Figure 1), the change in group similarity (Figure 3) and the variability in NSM (figures 2 and 4).

Moreover, figures 2 and 4 show that when the absolute value of the difference between $a$ and $b$ was large, almost all markers were selected (NSM was close to the total number of loci) suggesting an opposite behavior to the one described in theorems MC and MPC. This leads to the need for investigating the scenario of subject specific knowledge permitting to approximate the number of markers influenced by selection (a specific value or an interval) and looking for a subjective prior matching this information.

# 6. Final remarks

This study provides theoretical and empirical evidence on the asymptotic behaviour of a method used to infer selection signatures through genomic data that can be found when adopting a subjective Bayesian modelling approach. It was found that, the set of genomic regions flagged as potentially influenced by genetic selection change when varying the prior distribution parameters.

Sensitivity to hyperparameters is an important issue in any Bayesian analysis, besides, a mathematical approach to find the properties of an inferential method is scientifically more rigorous than those based on empirical analyses. Consequently, the main contribution of this study is the analytical approach to evaluate the impact of hyperparameters on this problem. Albeit the mathematical intractability of the posterior distribution of $\boldsymbol{\theta}$, the parameter of major interest, theorems MC and MPC uncover useful limiting properties, thus, they are regarded as our more significant contributions.

# Authors' contributions

Carlos Martínez: Conceptualization, Data curation, Formal analysis, Funding acquisition, Methodology, Writing - original draft, proofs of all formal results.
Nelson Cruz: Formal analysis, Methodology, Writing - review & editing, proofs of theorems 1 and 2.
Ronnal Ortiz: Data curation, Validation, Writing - review & editing.
Carmen Cepeda: Formal analysis, Validation, Writing - review & editing.
Juan Sosa: Data curation, Validation, Writing - review & editing.

**Proofs of theorems MC and MPC**

For the sake of clarity, before writing the proofs we present some results that will be used across this file.

**Theorem 1.** (Shao, 2007). Let $X, X_1, X_2, \ldots$be k-dimensional random vectors and $g\colon \mathbb{R}^k \longrightarrow \mathbb{R}^l$ be continuous a.s., then:

$$\{X_n\}_{n\geq 1} \xrightarrow{a.s.} X \text{ implies } g(X_n) \xrightarrow{a.s.} g(X)$$
$$\{X_n\}_{n\geq 1} \xrightarrow{P} X \text{ implies } g(X_n) \xrightarrow{P} g(X)$$
$$\{X_n\}_{n\geq 1} \xrightarrow{D} X \text{ implies } g(X_n) \xrightarrow{D} g(X)$$

where $a.\ s.$ stands for convergence almost surely, $P$ indicates converge in probability and $D$ convergence in distribution.

**Theorem 2**. (Casella and Berger, 2002). Let $n, m \in \mathbb{Z}_+$, $m > n$, $\{X_i\}_{i=1}^{m}$ be iid random variables with $X_i \sim Exp(\beta)$, $Y = \sum_{i=1}^{n} X_i$, and $Z = \sum_{i=1}^{m} X_i$, then $\frac{Y}{Z} \sim Beta(n, m-n)$.

**Theorem 3**. (The Moore-Osgood Theorem). Let $X$ be a metric space, $E$ a subspace of $X$, and $x_0$ an accumulation point of $E$. Let $f_n\colon X \to \mathbb{R}\ \forall\ n \in \mathbb{N}$ and $f\colon X \to \mathbb{R}$ be functions and $A_n$ numbers. If $f_n$ converges to $f$ uniformly on $E$ and $\lim_{x\to x_0} f_n(x) = A_n\ \forall\ n \in \mathbb{N}$, then the double limit $\lim_{(x,n)\to(x_0,\infty)} f_n(x)$ and the iterated limits $\lim_{n\to\infty}\lim_{x\to x_0} f_n(x)$ and $\lim_{x\to x_0}\lim_{n\to\infty} f_n(x)$, exist and they are equal, that is,

$$\lim_{(x,n)\to(x_0,\infty)} f_n(x) = \lim_{n\to\infty}\lim_{x\to x_0} f_n(x) = \lim_{x\to x_0}\lim_{n\to\infty} f_n(x)$$

This is a version of the Moore-Osgood theorem that can be found in Bartle (1991). The general form of this theorem can be found in Taylor (1985).

**Legendre duplication formula**. Let $\Gamma(\cdot)$ be the Gamma function and $x > 0$, then:

$$\Gamma(x)\Gamma\left(x + \frac{1}{2}\right) = 2^{1-2x}\sqrt{\pi}\,\Gamma(2x).$$

**Theorem 4**. (From Theorem 2 in Jameson, 2013). For $x > 0$ and $0 \leq y \leq 1$

$$x(x+y)^{y-1} \leq \frac{\Gamma(x+y)}{\Gamma(x)} \leq x^y.$$

**Theorem 5**. (Rudin, 1976). Let $D \subseteq \mathbb{R}$, unbounded above, be the domain of a function $f$, then $\lim_{x\to\infty} f(x) = L$, if and only if, for every sequence $\{x_n\}$ such that $\lim_{n\to\infty} x_n = \infty$, the sequence $\{f(x_n)\}$ converges to $L$.

**Proof of the MC Theorem.**

The rule to decide if $\hat{\theta}_l, l = 1,2, \ldots, L$, departs from what would be expected under the null model in the G-method, is based on checking if the posterior mean under the full model falls within a credibility set (interval) computed from the null posterior distribution. To this end, set $\alpha > 0$, small, and then, for $l = 1,2, \ldots, L$, declare $\hat{\theta}_l$ as departing from the null model if

$$\hat{\theta}_l \notin [\wp_{\alpha/2}, \wp_{1-\alpha/2}]$$

where $\wp_{\alpha/2}$ and $\wp_{1-\alpha/2}$ are the $\frac{\alpha}{2}$ and $1 - \frac{\alpha}{2}$ quantiles of the null posterior distribution of $\theta_l$, respectively. Notice that, the more similar the null and full posterior distributions of $\theta_l$ are, the less the number of markers with extreme $\hat{\theta}_l$ values under the null model. Recall that, a posteriori, $\{P_{lr}\}_{l,r=1}^{L,R}$ follow independent $Beta(n_{Alr} + a, n_{Blr} + b)$ distributions and the restriction $n_{Alr} + n_{Blr} = 2n_r$. Let $a = b = \delta$, and $\boldsymbol{x}$ be the observed data then

$$f(p_{lr}|\boldsymbol{x}, \delta) = \frac{p_{lr}^{n_{Alr}+\delta-1}(1-p_{lr})^{n_{Blr}+\delta-1}}{\frac{(\Gamma(\delta))^2}{\Gamma(2\delta)}}$$
$$= p_{lr}^{n_{Alr}-1}(1-p_{lr})^{n_{Blr}-1}[p_{lr}(1-p_{lr})]^{\delta} \frac{\Gamma(2\delta)}{(\Gamma(\delta))^2}$$

applying the Legendre duplication formula, we get:

$$\frac{\Gamma(2\delta)}{(\Gamma(\delta))^2} = \frac{\Gamma\left(\delta + \frac{1}{2}\right) 2^{2\delta}}{2\Gamma(\delta)\sqrt{\pi}}$$

thus,

$$f(p_{lr}|\boldsymbol{x}, \delta) = p_{lr}^{n_{Alr}-1}(1-p_{lr})^{n_{Blr}-1}[p_{lr}(1-p_{lr})]^{\delta} \frac{\Gamma\left(\delta + \frac{1}{2}\right) 2^{2\delta}}{2\Gamma(\delta)\sqrt{\pi}}$$
$$= p_{lr}^{n_{Alr}-1}(1-p_{lr})^{n_{Blr}-1}[4p_{lr}(1-p_{lr})]^{\delta} \frac{1}{2\sqrt{\pi}} \frac{\Gamma\left(\delta + \frac{1}{2}\right)}{\Gamma(\delta)}$$

thus, by Theorem 4 it follows that,

$$0 \leq f(p_{lr}|\boldsymbol{x},\delta) \leq p_{lr}^{n_{Alr}-1}(1-p_{lr})^{n_{Blr}-1}[4p_{lr}(1-p_{lr})]^{\delta}\,\frac{\delta^{\frac{1}{2}}}{2\sqrt{\pi}}$$

note that $\lim_{\delta\to\infty}[4p_{lr}(1-p_{lr})]^{\delta}\delta^{\frac{1}{2}} = 0$, iff $4p_{lr}(1-p_{lr}) < 1$, which amounts to $p_{lr}(1-p_{lr}) < \frac{1}{4}$, and this is equivalent to $p_{lr} \neq \frac{1}{2}$ because the function $p_{lr}(1-p_{lr})$ is maximized at $p_{lr} = \frac{1}{2}$ where it takes the value $\frac{1}{4}$. Consequently, for $p_{lr} \neq \frac{1}{2}$, by the squeeze theorem (Kosmala, 2004):

$$\lim_{\delta\to\infty} f(p_{lr}|\boldsymbol{x},\delta) = 0$$

Now this fact is used to prove convergence in probability. Let $P(\cdot)$ represent a probability measure. For arbitrary and small $\varepsilon > 0$

$$\lim_{\delta\to\infty} P(|P_{lr}-1/2| > \varepsilon|\boldsymbol{x},\delta) = \lim_{\delta\to\infty}\left[\int_{0}^{\frac{1}{2}-\varepsilon} f(p_{lr}|\boldsymbol{x},\delta)dp_{lr} + \int_{\frac{1}{2}+\varepsilon}^{1} f(p_{lr}|\boldsymbol{x},\delta)dp_{lr}\right]$$

$$= \lim_{\delta\to\infty}\int_{0}^{\frac{1}{2}-\varepsilon} f(p_{lr}|\boldsymbol{x},\delta)dp_{lr} + \lim_{\delta\to\infty}\int_{\frac{1}{2}+\varepsilon}^{1} f(p_{lr}|\boldsymbol{x},\delta)dp_{lr}$$

since we are dealing with a member of the exponential family of distributions, the limit and the integral can be exchanged (Shao, 2007) and the last expression is equal to

$$\int_{0}^{\frac{1}{2}-\varepsilon} \lim_{\delta\to\infty} f(p_{lr}|\boldsymbol{x},\delta)dp_{lr} + \int_{\frac{1}{2}+\varepsilon}^{1} \lim_{\delta\to\infty} f(p_{lr}|\boldsymbol{x},\delta)dp_{lr}$$

but we showed that for $p_{lr} \neq \frac{1}{2}$, $\lim_{\delta\to\infty} f(p_{lr}|\boldsymbol{x},\delta) = 0$, then

$$\int_{0}^{\frac{1}{2}-\varepsilon} \lim_{\delta\to\infty} f(p_{lr}|\boldsymbol{x},\delta)dp_{lr} + \int_{\frac{1}{2}+\varepsilon}^{1} \lim_{\delta\to\infty} f(p_{lr}|\boldsymbol{x},\delta)dp_{lr} = 0$$

In summary, we showed that for arbitrary and small $\varepsilon > 0, \forall\ l = 1,2,\ldots,L;\ \forall\ r = 1,2,\ldots,R$:

$$\lim_{\delta\to\infty} P(|P_{lr} - 1/2| > \varepsilon|\boldsymbol{x},\delta) = 0$$

i.e.,

$$P_{lr}|\boldsymbol{x},\delta \xrightarrow{\delta\to\infty} \frac{1}{2}\ in\ probability\ \forall\ l = 1,2,\dots,L;\ \forall\ r = 1,2,\dots,R.$$

Since for each $l$, the null posterior distribution is a $Beta(\sum_{r=1}^{R} n_{Alr} + \delta, \sum_{r=1}^{R} n_{Blr} + \delta)$ distribution, by similar arguments to those presented above, it follows that for all $l$, under this distribution $P_l$ converges in probability to $1/2$. Moreover, for each locus, the $F_{ST}$ parameter has the form:

$$\theta_l = \frac{\sum_{r=1}^{R} p_{lr}^2 - \frac{(\sum_{r=1}^{R} p_{lr})^2}{R}}{\sum_{r=1}^{R} p_{lr} - \frac{(\sum_{r=1}^{R} p_{lr})^2}{R}}$$

a real-valued continuous function of the random vector $(P_{l1}, P_{l2}, \dots, P_{lR})$ which takes the value of zero when $p_{l1} = p_{l1} = \cdots = p_{lR}$; hence, by Theorem 1

$$\theta_l|\boldsymbol{x},\delta \xrightarrow{\delta\to\infty} 0\ in\ probability\ \forall\ l = 1,2,\dots,L.$$

In addition, this result also holds under the null model and consequently, in view of the fact that convergence in probability implies convergence in distribution (Resnick, 2003), in the limit, the posterior distribution of $\theta_l$ is the same under both, the full and the null model and, as a result, for $\delta$ large enough, no extreme values of $\hat{\theta}_l$ are observed. This proves the desired result ■

However, we consider that it is worth pointing out and proving an interesting result for the special case of $\delta$ being a positive integer. Under this setting, a stronger sort of convergence known as almost surely or almost everywhere convergence, can be stablished. To this end, let $\delta \in \mathbb{Z}_+$ then, by Theorem 2 it follows that $\forall\ l = 1,2,\dots,L;\ \forall\ r = 1,2,\dots,R$:

$$\frac{\sum_{i=1}^{n_{Alr}+\delta} Y_i}{\sum_{i=1}^{2(n_r+\delta)} Y_i} \sim Beta(n_{Aj} + \delta, n_{Bj} + \delta)$$

with $\{Y_i\}_{i=1}^{2(n_j+\delta)}$ distributed iid $Exp(\beta)$. Now write

$$\frac{\sum_{i=1}^{n_{Alr}+\delta} Y_i}{\sum_{i=1}^{2(n_r+\delta)} Y_i} = \frac{\frac{1}{n_{Alr}+\delta}\sum_{i=1}^{n_{Alr}+\delta} Y_i}{\frac{2(n_r+\delta)}{n_{Alr}+\delta}\left(\frac{1}{2(n_r+\delta)}\sum_{i=1}^{2(n_r+\delta)} Y_i\right)}$$

Letting $\delta \to \infty$, then $n_{Alr} + \delta \to \infty$ and also $2n_r + \delta \to \infty$, thus, by properties of limits, the strong law or large numbers and Theorem 1, it follows that:

$$\frac{\frac{1}{n_{Alr}+\delta}\sum_{i=1}^{n_{Alr}+\delta} Y_i}{\frac{2(n_r+\delta)}{n_{Alr}+\delta}\left(\frac{1}{2(n_r+\delta)}\sum_{i=1}^{2(n_r+\delta)} Y_i\right)} \xrightarrow{\delta\to\infty} \frac{1}{2}\ a.s.$$

where $a.s.$ stands for almost surely. This result also holds under the null model and; therefore, applying Theorem 1 again, in the limit, the posterior distribution of $\theta_l$ is the same under both models which leads to the result stated in the theorem ■

**Proof of the MPC Theorem.**

Let $\boldsymbol{x}$ be the observed data and $\Delta \coloneqq a - b$. Following a procedure similar to the one used in the proof of the MC Theorem, it is shown that nder both, the full and the null model

$$\theta_l | \boldsymbol{x}, \Delta, b \xrightarrow{(\Delta,b)\to(0,\infty)} 0\ in\ distribution\ \forall\ l = 1,2,\dots,L$$

and consequently, in the limit, there are no markers with extreme $F_{ST}$ values.

Standard asymptotic theory cannot be used here since we are evaluating a double limit. Firstly, we use the density function of allele frequencies and show that the double limit is equal to zero when $p \neq 1/2$. Upon reparameterization in terms of $\Delta$, and $b \in \mathbb{N}$, the posterior pdf of $P_{lr}$ for arbitrary $l$ and $r$; $l = 1,2,\dots,L;\ r = 1,2,\dots,R$, is

$$\begin{aligned} f(p_{lr}|\boldsymbol{x},\Delta,b) &= \frac{p_{lr}^{n_{Alr}+\Delta+b-1}(1-p_{lr})^{n_{Blr}+b-1}}{\frac{\Gamma(\Delta+b)\Gamma(b)}{\Gamma(\Delta+2b)}} \\ &= p_{lr}^{n_{Alr}+b-1}(1-p_{lr})^{n_{Blr}+b-1}p_{lr}^{\Delta}\frac{\Gamma(\Delta+2b)}{\Gamma(\Delta+b)\Gamma(b)} \\ &= p_{lr}^{n_{Alr}-1}(1-p_{lr})^{n_{Blr}-1}[p_{lr}(1-p_{lr})]^b p_{lr}^{\Delta}\frac{\Gamma(\Delta+2b)}{\Gamma(\Delta+b)\Gamma(b)} \end{aligned}$$

Let

$$f_b(\Delta) \coloneqq [p_{lr}(1-p_{lr})]^b p_{lr}^{\Delta}\frac{\Gamma(\Delta+2b)}{\Gamma(\Delta+b)\Gamma(b)}$$

Now, since we are interested in the behavior of the function when $\Delta$ tends to 0, hereinafter we will focus on a neighborhood of 0, in particular we consider $\Delta \in [-1,1]$. In this interval, $\frac{\Gamma(\Delta+2b)}{\Gamma(\Delta+b)\Gamma(b)}$ is increasing in $\Delta$, consequently, $\forall\ \Delta \in [-1,1]$:

$$\frac{\Gamma(\Delta+2b)}{\Gamma(\Delta+b)\Gamma(b)} \leq \frac{\Gamma(1+2b)}{\Gamma(1+b)\Gamma(b)} \tag{1}$$

Applying the property $\Gamma(x+1) = x\Gamma(x), x > 0$ and the Legendre duplication formula, it follows that

$$\begin{aligned}\frac{\Gamma(1+2b)}{\Gamma(1+b)} &= \frac{2b\Gamma(2b)}{b\Gamma(b)} \\ &= \frac{2\Gamma(2b)}{\Gamma(b)} \\ &= \frac{2^{2b}\Gamma\left(\frac{1}{2}+b\right)}{\sqrt{\pi}}\end{aligned} \tag{2}$$

by (1), (2), Theorem 4, the fact that $\max_{\Delta\in[-1,1]}\left(p_{lr}^{\Delta}\right) = p_{lr}^{-1}$ and letting $\alpha_{p_{lr}} \coloneqq p_{lr}(1-p_{lr})$ it follows that

$$\begin{aligned}0 \leq f_b(\Delta) &\leq \left(\alpha_{p_{lr}}\right)^b p_{lr}^{\Delta} \frac{2^{2b}\Gamma\left(\frac{1}{2}+b\right)}{\sqrt{\pi}\Gamma(b)} \\ &\leq \left(\alpha_{p_{lr}}\right)^b \frac{2^{2b}\Gamma\left(\frac{1}{2}+b\right)}{p_{lr}\sqrt{\pi}\Gamma(b)} \\ &= \left(4\alpha_{p_{lr}}\right)^b \frac{\Gamma\left(\frac{1}{2}+b\right)}{p_{lr}\sqrt{\pi}\Gamma(b)} \\ &\leq \frac{\left(4\alpha_{p_{lr}}\right)^b b^{\frac{1}{2}}}{p_{lr}\sqrt{\pi}}\end{aligned} \tag{3}$$

Notice that Theorem 4 is applied in the last inequality. Now, like in the proof of the MC Theorem, note that the limit of (3) as $b\to\infty$ is equal to zero iff $4\alpha_{p_{lr}} < 1$, that is, if $p_{lr}(1-p_{lr}) < 1/4$ which is equivalent to $p_{lr} \neq 1/2$. Therefore, when $p_{lr} \neq 1/2$, by the Squeeze Theorem (Kosmala, 2004),

$$\lim_{b\to\infty} f_b(\Delta) = 0\ \forall\ \Delta \in [-1,1]$$

which stablishes convergence to the function $f = 0\ \forall\ \Delta \in [-1,1]$. Also, notice that since $b > 0$, $f_b$ is continuous and, therefore, for each $b$

$$\lim_{\Delta\to 0} f_b(\Delta) = \lim_{\Delta\to 0}\left([p_{lr}(1-p_{lr})]^b p_{lr}^{\Delta} \frac{\Gamma(\Delta+2b)}{\Gamma(\Delta+b)\Gamma(b)}\right)$$
$$= [p_{lr}(1-p_{lr})]^b \frac{\Gamma(2b)}{\left(\Gamma(b)\right)^2}$$

Therefore, for $p_{lr} \neq 1/2$, the Moore-Osgood theorem can be applied obtaining

$$\lim_{(\Delta,b)\to(0,\infty)} f(p_{lr}|\boldsymbol{x},\Delta,b) = \lim_{\Delta\to 0}\lim_{b\to\infty} f(p_{lr}|\boldsymbol{x},\Delta,b) = \lim_{b\to\infty}\lim_{\Delta\to 0} f(p_{lr}|\boldsymbol{x},\Delta,b)$$

Now, since the posterior density is continuous as well, for each $b$

$$\lim_{\Delta\to 0} f(p_{lr}|\boldsymbol{x},\Delta,b) = p_{lr}^{n_{Alr}-1}(1-p_{lr})^{n_{Blr}-1}[p_{lr}(1-p_{lr})]^b \frac{\Gamma(2b)}{\left(\Gamma(b)\right)^2}$$

Applying the Legendre duplication formula again we get

$$\frac{\Gamma(2b)}{\left(\Gamma(b)\right)^2} = \frac{\Gamma\left(b+\frac{1}{2}\right)2^{2b}}{2\Gamma(b)\sqrt{\pi}}$$

thus, using this identity and Theorem 4:

$$[p_{lr}(1-p_{lr})]^b \frac{\Gamma(2b)}{\left(\Gamma(b)\right)^2} = \frac{[4p_{lr}(1-p_{lr})]^b}{2\sqrt{\pi}} \frac{\Gamma\left(b+\frac{1}{2}\right)}{\Gamma(b)}$$
$$\leq \frac{[4p_{lr}(1-p_{lr})]^b}{2\sqrt{\pi}} b^{1/2}$$

also, for $p_{lr} \neq 1/2$

$$\lim_{b\to\infty} \frac{[4p_{lr}(1-p_{lr})]^b}{2\sqrt{\pi}} b^{1/2} = 0$$

thus, applying the squeeze theorem and the Moore-Osgood theorem

$$\lim_{b\to\infty}\lim_{\Delta\to 0} f(p_{lr}|\boldsymbol{x},\Delta,b) = 0$$
$$= \lim_{(\Delta,b)\to(0,\infty)} f(p_{lr}|\boldsymbol{x},\Delta,b)$$

The extension to the case $b \in \mathbb{R}_+$, the set of positive real numbers, is straightforward by replacing $b$ in the derivations above by an arbitrary sequence $\{w_n\}_{n\in\mathbb{N}}$ growing to infinity. Then

$$[p_{lr}(1-p_{lr})]^{w_n} p_{lr}^{\Delta} \frac{\Gamma(\Delta+2w_n)}{\Gamma(\Delta+w_n)\Gamma(w_n)}$$

is a function that plays the role of $f_b(\Delta)$ above becase it depends on $n \in \mathbb{N}$ and tends to infinity as $n$ tends to infinity. The rest of the procedure is the same and, therefore, we find out that the double limit exists, and it is equal to the iterated limits and, consequently, after taking the limit of $f(p_{lr}|\boldsymbol{x},\Delta,b)$ as $\Delta$ tends to zero we get the function

$$\frac{[4p_{lr}(1-p_{lr})]^{w_n}}{2\sqrt{\pi}}(w_n)^{1/2}$$

and, for $p_{lr} \neq 1/2$

$$\lim_{n\to\infty} \frac{[4p_{lr}(1-p_{lr})]^{w_n}}{2\sqrt{\pi}}(w_n)^{1/2} = 0$$

since $w_n$ was arbitrary, by Theorem 5, we have convergence for $b \in \mathbb{R}_+$.

The remaining part of the proof is very similar to the one presented abovefor Theorem MC, but for the sake of clarity and completness, the details are presented bellow.

Let $P(\cdot)$ represent a probability measure. For arbitrary and small $\varepsilon > 0, \forall\, l = 1,2,\dots,L;\ r = 1,2,\dots,R$

$$\lim_{(\Delta,b)\to(0,\infty)} P(|P_{lr}-1/2| > \varepsilon|\boldsymbol{x},\Delta,b) = \lim_{(\Delta,b)\to(0,\infty)} \left[\int_0^{\frac{1}{2}-\varepsilon} f(p_{lr}|\boldsymbol{x},\delta)dp_{lr} + \int_{\frac{1}{2}+\varepsilon}^{1} f(p_{lr}|\boldsymbol{x},\delta)dp_{lr}\right]$$

$$= \lim_{(\Delta,b)\to(0,\infty)} \int_0^{\frac{1}{2}-\varepsilon} f(p_{lr}|\boldsymbol{x},\delta)dp_{lr} + \lim_{(\Delta,b)\to(0,\infty)} \int_{\frac{1}{2}+\varepsilon}^{1} f(p_{lr}|\boldsymbol{x},\delta)dp_{lr}$$

using the fact that we are dealing with a member of the exponential family of distributions again, the last expression is equal to

$$\int_{0}^{\frac{1}{2}-\varepsilon} \lim_{(\Delta,b)\to(0,\infty)} f(p_{lr}|\boldsymbol{x},\delta)dp_{lr} + \int_{\frac{1}{2}+\varepsilon}^{1} \lim_{(\Delta,b)\to(0,\infty)} f(p_{lr}|\boldsymbol{x},\delta)dp_{lr}$$

but we already proved that for $p_{lr} \neq 1/2$ ; $\lim_{(\Delta,b)\to(0,\infty)} f(p_{lr}|\boldsymbol{x},\delta) = 0$ so both integrands are null. Notice that we just did show that for arbitrary and small $\varepsilon > 0$

$$\lim_{(\Delta,b)\to(0,\infty)} P(|P_{lr} - 1/2| > \varepsilon|\boldsymbol{x},\Delta,b) = 0$$
$$\forall l = 1,2,\dots,L;\ r = 1,2,\dots,R$$

i.e.,

$$P_{lr}|\boldsymbol{x},\Delta,b \xrightarrow{(\Delta,b)\to(0,\infty)} \frac{1}{2}\ \forall l\ \forall r\ in\ probability$$

Thus, because convergence in probability implies convergence in distribution (Resnick, 2003) and by Theorem1 it follows that:

$$\theta_l|\boldsymbol{x},\Delta,b \xrightarrow{(\Delta,b)\to(0,\infty)} 0\ \forall l\ in\ distribution$$

Finally, following the same rationale used in the proof of the MC Theorem, in the limit the null and full posterior distributions of $\theta_l$ degenerate to 0 $\forall\ l = 1,2,\dots,L$, which proves the theorem ■

## Mathematical derivations associated to the posterior mean and variance functions of allele frequencies

**Derivation of the gradient vector and Hessian matrix of the posterior mean function of allele frequencies.**

Recall that:

$$\mu_{P_{lr}}^{Post}(a,b) = \frac{n_{Arl} + a}{2n_r + a + b}$$
$$a, b > 0;\ n_{Arl}, n_{Arl} \in \mathbb{Z}_*;\ 2n_r = n_{Arl} + n_{Brl}$$

where $\mathbb{Z}_*$ is the set of non-negative integers.

Using the quotient rule we get:

$$\frac{\partial \mu_{P_{lr}}^{Post}}{\partial a} = \frac{2n_r + a + b - (n_{Arl} + a)}{(2n_r + a + b)^2} = \frac{n_{Brl} + b}{(2n_r + a + b)^2}$$

and proceeding similarly:

$$\frac{\partial \mu_{P_{lr}}^{Post}}{\partial b} = -\frac{n_{Arl} + a}{(2n_r + a + b)^2}$$

Now, using the quotient rule again we find the second partial derivatives

$$\frac{\partial^2 \mu_{P_{lr}}^{Post}}{\partial a^2} = -\frac{2(2n_r + a + b)(n_{Brl} + b)}{2(2n_r + a + b)^4}$$
$$= -\frac{2(n_{Brl} + b)}{(2n_r + a + b)^3}$$
$$\frac{\partial^2 \mu_{P_{lr}}^{Post}}{\partial b^2} = \frac{2(2n_r + a + b)(n_{Arl} + a)}{(2n_r + a + b)^4}$$
$$= \frac{2(n_{Arl} + a)}{(2n_r + a + b)^3}$$

and the mixed partial derivatives

$$\frac{\partial^2 \mu_{P_{lr}}^{Post}}{\partial a \partial b} = \frac{(2n_r + a + b)^2 - 2(2n_r + a + b)(n_{arl} + b)}{(2n_r + a + b)^4}$$
$$= \frac{(2n_r + a + b)[2n_r + a + b - 2n_{arl} - 2b]}{(2n_r + a + b)^4}$$
$$\frac{n_{Arl} - n_{Brl} + a - b}{(2n_r + a + b)^3} = \frac{\partial^2 \mu_{P_{lr}}^{Post}}{\partial b \partial a}$$

**Derivation of gradient vector and Hessian matrix of the posterior variance function of allele frequencies.**

Recall that

$$V_{P_{lr}}^{Post}(a,b) = \frac{(n_{Arl}+a)(n_{Brl}+b)}{(2n_r+a+b)^2(2n_r+a+b+1)}$$

**Partial derivatives**

Using the quotient rule, the product rule and $2n_r = n_{Arl} + n_{Brl}$ we get

$$\frac{\partial V_{P_{lr}}^{Post}}{\partial a} = \frac{(n_{Brl}+b)(2n_r+a+b)^2(2n_r+a+b+1) - [2(2n_r+a+b)(2n_r+a+b+1)+(2n_r+a+b)^2](n_{Arl}+a)(n_{Brl}+b)}{(2n_r+a+b)^4(2n_r+a+b+1)^2}$$
$$= \frac{(n_{Brl}+b)(2n_r+a+b)[(2n_r+a+b)(2n_r+a+b+1) - 2(2n_r+a+b+1)(n_{Arl}+a) - (2n_r+a+b)(n_{Arl}+a)]}{(2n_r+a+b)^4(2n_r+a+b+1)^2}$$
$$= \frac{(n_{Brl}+b)[(2n_r+a+b+1)(2n_r+a+b-2n_{Arl}-2a) - (2n_r+a+b)(n_{Arl}+a)]}{(2n_r+a+b)^3(2n_r+a+b+1)^2}$$
$$= \frac{(n_{Brl}+b)[(2n_r+a+b+1)(n_{Brl}+b-n_{Arl}-a) - (2n_r+a+b)(n_{Arl}+a)]}{(2n_r+a+b)^3(2n_r+a+b+1)^2}$$
$$= \frac{(n_{Brl}+b)[(2n_r+a+b+1)[n_{Brl}-n_{Arl}+b-a] - (2n_r+a+b)(n_{Arl}+a)]}{(2n_r+a+b)^3(2n_r+a+b+1)^2}$$

$$\frac{\partial V_{P_{lr}}^{Post}}{\partial b} = \frac{(n_{Arl}+a)(2n_r+a+b)^2(2n_r+a+b+1) - [2(2n_r+a+b)(2n_r+a+b+1)+(2n_r+a+b)^2](n_{Arl}+a)(n_{Brl}+b)}{(2n_r+a+b)^4(2n_r+a+b+1)^2}$$
$$= \frac{(n_{Arl}+a)(2n_r+a+b)[(2n_r+a+b)(2n_r+a+b+1) - 2(2n_r+a+b+1)(n_{Brl}+b) - (2n_r+a+b)(n_{Brl}+b)]}{(2n_r+a+b)^4(2n_r+a+b+1)^2}$$
$$= \frac{(n_{Arl}+a)[(2n_r+a+b+1)(2n_r+a+b-2n_{Brl}-2b) - (2n_r+a+b)(n_{Brl}+b)]}{(2n_r+a+b)^3(2n_r+a+b+1)^2}$$

$$= \frac{(n_{Arl} + a)[(2n_r + a + b + 1)(n_{Arl} + a - n_{Brl} - b) - (2n_r + a + b)(n_{Brl} + b)]}{(2n_r + a + b)^3(2n_r + a + b + 1)^2}$$
$$= \frac{(n_{Arl} + a)[(2n_r + a + b + 1)[n_{Arl} - n_{Brl} + a - b] - (2n_r + a + b)(n_{Brl} + b)]}{(2n_r + a + b)^3(2n_r + a + b + 1)^2}$$

**Second partial derivatives.**

Let $v_{a_1} = (n_{Blr} + b)[(2n_r + a + b + 1)(n_{Blr} - n_{Alr} + b - a) - (2n_r + a + b)(n_{Alr} + a)]$ and $v_2 = (2n_r + a + b)^3(2n_r + a + b + 1)^2$, then

$$\frac{\partial^2 V_{P_{lr}}^{Post}}{\partial a^2} = \frac{(n_{Brl} - n_{Arl} + b - a - 2n_r - a - b - 1 - n_{Arl} - a - 2n_r - a - b)(n_{Brl} + b)v_2 - [3(2n_r + a + b)^2(2n_r + a + b + 1)^2 + 2(2n_r + a + b + 1)(2n_r + a + b)^3](n_{Brl} + b)[(2n_r + a + b + 1)v_{a_1}}{(2n_r + a + b)^6(2n_r + a + b + 1)^4}$$
$$= \frac{(n_{Brl} - 2n_{Arl} - b - 4a - 4n_r - 1)(n_{Brl} + b)(2n_r + a + b)^3(2n_r + a + b + 1)^2 - [3(2n_r + a + b)^2(2n_r + a + b + 1)^2 + 2(2n_r + a + b + 1)(2n_r + a + b)^3]v_{a_1}}{(2n_r + a + b)^6(2n_r + a + b + 1)^4}$$
$$= \frac{(2n_r + a + b)^2(2n_r + a + b + 1)\left[((n_{Brl} + b)((n_{Brl} - 2(n_{Arl} - 4n_r - b - 1)(2n_r + a + b)(2n_r + a + b + 1) - \left(3(2n_r + a + b + 1) + 2(2n_r + a + b)\right)v_{a_1}\right]}{(2n_r + a + b)^6(2n_r + a + b + 1)^4}$$
$$= \frac{(n_{Brl} + b)(n_{Brl} - 2(n_{Arl} - 4n_r - 4a - b - 1)(2n_r + a + b)(2n_r + a + b + 1) - (6n_r + 3a + 3b + 3 + 4n_r + 2a + 2b)v_{a_1}}{(2n_r + a + b)^4(2n_r + a + b + 1)^3}$$
$$= \frac{(n_{Brl} + b)\left[(n_{Brl} - 2n_{Arl} - 4(n_r + a) - b - 1)(2n_r + a + b)(2n_r + a + b + 1) - (10n_r + 5(a + b) + 3)[(2n_r + a + b + 1)[n_{Brl} - n_{Arl} + b - a] - (2n_r + a + b)(n_{Arl} + a)]\right]}{(2n_r + a + b)^4(2n_r + a + b + 1)^3}$$

Now, let $v_{b_1} = (n_{Arl} + a)[(2n_r + a + b + 1)(n_{Arl} - n_{Brl} + a - b) - (2n_r + a + b)(n_{Brl} + b)]$, then

$$\frac{\partial^2 V_{P_{lr}}^{Post}}{\partial b^2} = \frac{(n_{Arl} - n_{Brl} + a - b - 2n_r - a - b - 1 - n_{Brl} - b - 2n_r - a - b)(n_{Arl} + a)v_2 - [3(2n_r + a + b)^2(2n_r + a + b + 1)^2 + 2(2n_r + a + b + 1)(2n_r + a + b)^3]v_{b_1}}{(2n_r + a + b)^6(2n_r + a + b + 1)^2}$$
$$= \frac{(n_{Arl} - 2n_{Brl} - a - 4b - 4n_r - 1)(n_{Arl} + a)(2n_r + a + b)^3(2n_r + a + b + 1)^2 - \left[(2n_r + a + b)^2(2n_r + a + b + 1)[6n_r + 3a + 3b + 3 + 4n_r + 2a + 2b]\right]v_{b_1}}{(2n_r + a + b)^6(2n_r + a + b + 1)^4}$$
$$= \frac{(2n_r + a + b)^2(2n_r + a + b + 1)\left[(n_{Arl} + a)(n_{Arl} - 2n_{Brl} - 4b - 4n_r - a - 1)(2n_r + a + b)(2n_r + a + b + 1) - (10n_r + 5a + 5b + 3)v_{b_1}\right]}{(2n_r + a + b)^6(2n_r + a + b + 1)^4}$$
$$= \frac{(n_{Arl} + a)(n_{Arl} - 2n_{Brl} - 4b - 4n_r - a - 1)(2n_r + a + b)(2n_r + a + b + 1) - (10n_r + 5a + 5b + 3)v_{b_1}}{(2n_r + a + b)^4(2n_r + a + b + 1)^3}$$

$$= \frac{(n_{Arl}+a)\big[(n_{Arl}-2n_{Brl}-4(n_r+b)-a-1)(2n_r+a+b)(2n_r+a+b+1)-(10n_r+5(a+b)+3)[(2n_r+a+b+1)(n_{Arl}-n_{Brl}+a-b)-(2n_r+a+b)(n_{Brl}+b)]\big]}{(2n_r+a+b)^4(2n_r+a+b+1)^3}$$

Proceeding similarly, the mixed partial derivatives are computed as follows. Let $v_{ab} = (n_{Arl}+a)[(2n_r+a+b+1)[n_{Arl}-n_{Brl}+a-b]-(2n_r+a+b)(n_{Brl}+b)]$, then

$$\frac{\partial^2 V_{P_{lr}}^{Post}}{\partial b \partial a} = \frac{[(2n_r+a+b+1)[n_{Arl}-n_{Brl}+a-b]-(2n_r+a+b)(n_{Brl}+b)-(n_{Arl}-n_{Brl}+a-b+2n_r+a+b+1-n_{Brl}-b-2n_r-a-b)(n_{Arl}+a)]v_2-[3(2n_r+a+b)^2(2n_r+a+b+1)^2+2(2n_r+a+b+1)(2n_r+a+b)^3]v_{ab}}{(2n_r+a+b)^6(2n_r+a+b+1)^4}$$

$$= \frac{(2n_r+a+b)^2(2n_r+a+b+1)\big[\big((2n_r+a+b+1)[n_{Arl}-n_{Brl}+a-b]-(2n_r+a+b)(n_{Brl}+b)+(n_{Arl}-2n_{Brl}+a-2b+1)(n_{Arl}+a)\big)(2n_r+a+b)(2n_r+a+b+1)-(6n_r+3a+3b+3+4n_r+2a+2b)v_{ab}\big]}{(2n_r+a+b)^6(2n_r+a+b+1)^4}$$

$$= \frac{[(2n_r+a+b+1)[n_{Arl}-n_{Brl}+a-b]-(2n_r+a+b)(n_{Brl}+b)][(2n_r+a+b)(2n_r+a+b+1)-(10n_r+5a+5b+3)(n_{Arl}+a)]+(n_{Arl}-2n_{Brl}+a-2b+1)(2n_r+a+b)(2n_r+a+b+1)(n_{Arl}+a)}{(2n_r+a+b)^4(2n_r+a+b+1)^3}$$

$$= \frac{[(2n_r+a+b+1)[n_{Arl}-n_{Brl}+a-b]-(2n_r+a+b)(n_{Brl}+b)][(2n_r+a+b)(2n_r+a+b+1)-(10n_r+5(a+b)+3)(n_{Arl}+a)]+(n_{Arl}-2n_{Brl}+a-2b+1)(2n_r+a+b)(2n_r+a+b+1)(n_{Arl}+a)}{(2n_r+a+b)^4(2n_r+a+b+1)^3}$$

$$= \frac{\partial^2 V_{P_{lr}}^{Post}}{\partial a \partial b}$$

**Proof of Proposition 1**.

Equating $\nabla_{2lr}$ to zero induces:

$$(2n_r + a + b + 1)[n_{Brl} - n_{Arl} + b - a] - (2n_r + a + b)(n_{Arl} + a) = 0$$
$$\Rightarrow (2n_r + a + b + 1)[n_{Brl} - n_{Arl} + b - a] = (2n_r + a + b)(n_{Arl} + a) \qquad (1)$$

and

$$(2n_r + a + b + 1)[n_{Arl} - n_{Brl} + a - b] - (2n_r + a + b)(n_{Brl} + b) = 0$$
$$\Rightarrow -(2n_r + a + b + 1)[n_{Brl} - n_{Arl} + b - a] - (2n_r + a + b)(n_{Brl} + b) = 0$$
$$\Rightarrow (2n_r + a + b + 1)[n_{Brl} - n_{Arl} + b - a] = -(2n_r + a + b)(n_{Brl} + b) \qquad (2)$$

since the left-hand sides of equations (1) and (2) are equal, it follows that

$$(2n_r + a + b)(n_{Arl} + a) = -(2n_r + a + b)(n_{Brl} + b)$$
$$\Rightarrow n_{Arl} + a = -n_{Brl} - b$$

which is a contradiction because $n_{Arl}, n_{arl}, a, b$ are all positive.

Now, from the partial derivative $\frac{\partial V_{P_{lr}}^{Post}}{\partial a}$, obtained above, it follows that for $b > 0$ arbitrary, but fixed, $V_{P_{lr}}^{Post}$ is strictly decreasing in $a$ iff

$$(2n_r + a + b + 1)(n_{Blr} - n_{Alr} + b - a) - (2n_r + a + b)(n_{Alr} + a) < 0$$

$$\Leftrightarrow \frac{n_{Blr} - n_{Alr} + b - a}{n_{Alr} + a} < \frac{2n_r + a + b}{2n_r + a + b + 1} \qquad (3)$$

Since $a, b, 2n_r, n_{Alr} > 0$, it is easy to see that if $n_{Blr} - n_{Alr} + b - a < 0$ then (3) is satisfied, but this is equivalent to $n_{Alr} + a > n_{Blr} + b$, that is, it holds in $\mathcal{D}_{lr1}$.
On the other hand, $V_{P_{lr}}^{Post}$ is increasing in $a$ for fixed $b$ iff

$$\frac{n_{Blr} - n_{Alr} + b - a}{n_{Alr} + a} > \frac{2n_r + a + b}{2n_r + a + b + 1}$$

Note that $n_{Blr} + b \geq 2(n_{Alr} + a)$ is equivalent to

$$\frac{n_{Blr} + b}{n_{Alr} + a} \geq 2 \Leftrightarrow \frac{n_{Blr} + b}{n_{Alr} + a} - 1 \geq 1$$
$$\Leftrightarrow \frac{n_{Blr} + b - (n_{Alr} + a)}{n_{Alr} + a} = \frac{n_{Blr} - n_{Alr} + b - a}{n_{Alr} + a} \geq 1 > \frac{2n_r + a + b}{2n_r + a + b + 1}$$

On the other hand, from the functional form of $\frac{\partial V_{P_{lr}}^{Post}}{\partial b}$, derived above, it follows that $V_{P_{lr}}^{Post}$ is strictly decreasing (increasing) in $b$ for fixed $a$ iff

$$(2n_r + a + b + 1)[n_{Arl} - n_{Brl} + a - b] - (2n_r + a + b)(n_{Brl} + b) < (>) 0$$

$$\Leftrightarrow \frac{n_{Arl} - n_{Brl} + a - b}{n_{Brl} + b} < (>) \frac{2n_r + a + b}{2n_r + a + b + 1}$$

So, the proof of the proposition for $\frac{\partial V_{P_{lr}}^{Post}}{\partial b}$ is completely analogous to the one for $\frac{\partial V_{P_{lr}}^{Post}}{\partial a}$, consequently, it is omitted. ■

**Proof of Proposition 2**.

From the derivations shown at the beginning of this file, the Hessian matrix of $V_{P_{lr}}^{Post}$ has the following form:

$$H_{2lr} = \frac{1}{D}\begin{bmatrix} N_a & N_{ab} \\ N_{ab} & N_b \end{bmatrix}$$

where

$$\begin{aligned} N_a = (n_{Blr} + b)[(n_{Blr} - 2n_{Alr} - b - 4(n_r + a) - 1)(2n_r + a + b)(2n_r + a + b + 1) + \\ -(10n_r + 5(a + b) + 3)] \times \\ \times [(2n_r + a + b + 1)(n_{Blr} - n_{Alr} + b - a) - (2n_r + a + b)(n_{Alr} + a)] \end{aligned}$$

$$\begin{aligned} N_b = (n_{Alr} + a)[(n_{Alr} - 2n_{Blr} - a - 4(n_r + b) - 1)(2n_r + a + b)(2n_r + a + b + 1) + \\ -(10n_r + 5(a + b) + 3)] \times \\ \times [(2n_r + a + b + 1)(n_{Alr} - n_{Blr} + a - b) - (2n_r + a + b)(n_{Blr} + b)] \end{aligned}$$

$$\begin{aligned} N_{ab} = [(2n_r + a + b + 1)(n_{Alr} - n_{Blr} + a - b) - (2n_r + a + b)(n_{Blr} + b)] \times \\ \times [(2n_r + a + b)(2n_r + a + b + 1) - (10n_r + 5(a + b) + 3)(n_{Alr} + a)] + \\ +(n_{Alr} - 2n_{Blr} + a - 2b + 1)(2n_r + a + b)(2n_r + a + b + 1)(n_{Alr} + a) \end{aligned}$$

$$D = (2n_r + a + b)^4(2n_r + a + b + 1)^3$$

Notice that $N_a$ is of the form $c(def - g)(eh - di)$, where

$$\begin{gathered} c = n_{Blr} + b \\ d = 2n_r + a + b \\ e = 2n_r + a + b + 1 = d + 1 \\ f = n_{Blr} - 2n_{Alr} - b - 4(a + n_r) - 1 \\ g = 10n_r + 5(a + b) + 3 \\ h = n_{Blr} - n_{Alr} + b - a \\ i = n_{Alr} + a \end{gathered}$$

hence $c, d, e, g, i > 0$. Now, let us analyze $f$. Recall the restriction $2n_r = n_{Alr} + n_{Blr}$, therefore, $n_{Blr} = 2n_r - n_{Alr}$, thus:

$$\begin{aligned} f &= 2n_r - n_{Alr} - 2n_{Alr} - 4n_r - (b + 4a + 1) \\ &= -2n_r - 3n_{Alr} - (b + 4a + 1) < 0 \;\forall\, a, b > 0 \end{aligned}$$

from this result and the fact that $c, d, e, g > 0$ it follows that

$$c(def - g) < 0 \;\forall\, a, b > 0 \tag{4}$$

from (4) and the definition of $N_a$

$$sign(N_a) = \begin{cases} -1, iff\ eh - di > 0 \\ 1, iff\ eh - di < 0 \end{cases}$$

$$\begin{gathered} eh - di < (>)0 \\ \Leftrightarrow (2n_r + a + b + 1)(n_{Blr} - n_{Alr} + b - a) < (>)(2n_r + a + b)(n_{Alr} + a) \\ \Leftrightarrow \frac{n_{Blr} - n_{Alr} + b - a}{n_{Alr} + a} < (>) \frac{2n_r + a + b}{2n_r + a + b + 1} \end{gathered}$$

$$\Leftrightarrow \frac{n_{Blr} + b}{n_{Alr} + a} < (>) \frac{2n_r + a + b}{2n_r + a + b + 1} + 1 \tag{5}$$

Similarly, $N_b$ is of the form $i(def^* - g)(eh^* - dc)$

$$f^* = n_{Alr} - 2n_{Blr} - a - 4(b + n_r) - 1$$
$$h^* = n_{Alr} - n_{Blr} + a - b = -h$$

using $n_{Alr} = 2n_r - n_{alr}$ and proceeding as done for $f$, it is easy to see that $f^* < 0$; therefore,

$$sign(N_b) = \begin{cases} -1, if\ eh^* - dc > 0 \\ 1, if\ eh^* - dc < 0 \end{cases}$$

$$eh^* - dc < (>)0$$
$$\Leftrightarrow \frac{n_{Alr} + a}{n_{Blr} + b} < (>) \frac{2n_r + a + b}{2n_r + a + b + 1} + 1 \quad (6)$$

Notice that

$$\frac{2n_r + a + b}{2n_r + a + b + 1} \in (0, 1)$$
$$\Leftrightarrow K_{x,a,b} = \frac{2n_r + a + b}{2n_r + a + b + 1} + 1 \in (1, 2)$$
$$\Leftrightarrow K_{x,a,b}^{-1} \in \left(\frac{1}{2}, 1\right)$$

hence

$$K_{x,a,b} > K_{x,a,b}^{-1} > 0$$

Thus, if $\frac{n_{Blr}+b}{n_{Alr}+a} > K_{x,a,b}$ (i.e., $N_a < 0$) then

$$\left(\frac{n_{Blr} + b}{n_{Alr} + a}\right)^{-1} = \frac{n_{Alr} + a}{n_{Blr} + b} < K_{x,a,b}^{-1} < K_{x,a,b}$$
$$\Rightarrow N_b > 0 \ \big(by\ (6)\big)$$

Similarly, if $\frac{n_{Alr}+a}{n_{Blr}+b} > K_{x,a,b}$ (i.e., $N_b < 0$) then

$$\left(\frac{n_{Alr} + a}{n_{Blr} + b}\right)^{-1} = \frac{n_{Blr} + b}{n_{Alr} + a} < K_{x,a,b}^{-1} < K_{x,a,b}$$
$$\Rightarrow N_a > 0 \ (by\ (5))$$

Moreover, if $\frac{n_{Alr}+a}{n_{Blr}+b} \in \left(K_{x,a,b}^{-1}, K_{x,a,b}\right)$, in particular $\frac{n_{Alr}+a}{n_{Blr}+b} > K_{x,a,b}^{-1} \Leftrightarrow \frac{n_{Blr}+b}{n_{Alr}+a} < K_{x,a,b}$, that is, $N_a > 0$. Also, by (6) it follows that

$$sign(N_b) = \begin{cases} 1, if\ \frac{n_{Alr}+a}{n_{Blr}+b} < K_{x,a,b} \\ -1, if\ \frac{n_{Alr}+a}{n_{Blr}+b} > K_{x,a,b} \end{cases}$$

and consequently, if $\frac{n_{Alr}+a}{n_{Blr}+b} \in \left(K_{x,a,b}^{-1}, K_{x,a,b}\right)$, then $N_b > 0$ and $N_a > 0$.

Analogously, by (5)

$$sign(N_a) = \begin{cases} 1, if\ \dfrac{n_{Blr} + b}{n_{Alr} + a} < K_{x,a,b} \\ -1, if\ \dfrac{n_{Blr} + b}{n_{Alr} + a} > K_{x,a,b} \end{cases}$$

so, if $\frac{n_{Blr}+b}{n_{Alr}+a} \in \left(K_{\boldsymbol{x},a,b}^{-1}, K_{\boldsymbol{x},a,b}\right)$ then $N_b > 0$ and $N_a > 0$

Because $H_{2lr}$ is symmetric and $\dim(H_{2lr}) = 2 \times 2$, whenever the signs of $N_a$ and $N_b$ are different $|H_{2lr}| < 0$, thus, the eigen values of $H_2$ have different signs (Harville, 1997) and therefore it is indefinite. Moreover, when any of the two conditions under which $N_a, N_b > 0$ are met, if $|H_{2lr}| > 0$ then the Hessian is positive definite and posterior variance function is strictly convex.

Now, consider the condition $N_a < 0$, it was already shown that this is equivalent to $\frac{n_{Blr}+b}{n_{Alr}+a} > K_{\boldsymbol{x},a,b}$ which implies that $\frac{n_{Alr}+a}{n_{alr}+b} < K_{\boldsymbol{x},a,b}^{-1} < K_{\boldsymbol{x},a,b}$ and $\frac{n_{Alr}+a}{n_{Blr}+b} < K_{\boldsymbol{x},a,b} \Leftrightarrow N_b > 0$ , similarly, $N_b < 0$ is equivalent to $\frac{n_{Alr}+a}{n_{Blr}+b} > K_{\boldsymbol{x},a,b}$ which implies that $\frac{n_{Blr}+b}{n_{Alr}+a} < K_{\boldsymbol{x},a,b}^{-1} < K_{\boldsymbol{x},a,b}$ and $\frac{n_{Blr}+b}{n_{Alr}+a} < K_{\boldsymbol{x},a,b} \Leftrightarrow N_a > 0$ which proves that the sets where the hyperparameters and the population structure satisfy $\frac{n_{Blr}+b}{n_{Alr}+a} > K_{\boldsymbol{x},a,b}$ and $\frac{n_{Alr}+a}{n_{Blr}+b} > K_{\boldsymbol{x},a,b}$ are disjoint and, consequently, $N_a$ and $N_b$ cannot be simultaneously negative. ■

**Mathematical derivations associated to the posterior predictive density of allele frequencies**

**Details on the derivation of the posterior predictive density.**

In the manuscript, it was shown that

$$m_\pi(\boldsymbol{x}|\boldsymbol{x}^{\sigma}) = NC\prod_{r=1}^{R}\prod_{l=1}^{L}\int_0^1 p_{lr}^{n_{Alr}+n_{Alr}^{\sigma}+a-1}(1-p_{lr})^{n_{Blr}+n_{Blr}^{\sigma}+b-1}\,dp_{lr}$$

where $NC = \prod_{r=1}^{R}\prod_{l=1}^{L}\frac{1}{B(n_{Alr}^{\sigma}+a,n_{Blr}^{\sigma}+b)}$. Since we are integrating the kernels of Beta densities it follows that

$$m_\pi(\boldsymbol{x}|\boldsymbol{x}^{\sigma}) = NC\prod_{r=1}^{R}\prod_{l=1}^{L}\frac{\Gamma(n_{Alr}+n_{Alr}^{\sigma}+a)\Gamma(n_{Blr}+n_{Blr}^{\sigma}+b)}{\Gamma(2(n_r+n_r^{\sigma})+a+b)}$$

using the definition of $NC$ (shown above) and the fact that the Beta function can be written as the ratio of Gamma functions we obtain

$$\begin{aligned} m_\pi(\boldsymbol{x}|\boldsymbol{x}^{\sigma}) &= \prod_{r=1}^{R}\prod_{l=1}^{L}\left(\frac{\Gamma(2n_r^{\sigma}+a+b)}{\Gamma(n_{Alr}^{\sigma}+a)\Gamma(n_{Blr}^{\sigma}+b)}\times\frac{\Gamma(n_{Alr}+n_{Alr}^{\sigma}+a)\Gamma(n_{Blr}+n_{Blr}^{\sigma}+b)}{\Gamma(2(n_r+n_r^{\sigma})+a+b)}\right) \\ &= \prod_{r=1}^{R}\prod_{l=1}^{L}\left(\frac{\Gamma(2n_r^{\sigma}+a+b)}{\Gamma(2(n_r+n_r^{\sigma})+a+b)}\times\frac{\Gamma(n_{Alr}+n_{Alr}^{\sigma}+a)}{\Gamma(n_{Alr}^{\sigma}+a)}\times\frac{\Gamma(n_{Blr}+n_{Blr}^{\sigma}+b)}{\Gamma(n_{Blr}^{\sigma}+b)}\right). \end{aligned}$$

Now, using the recurrence property of the Gamma function $\Gamma(x+1) = x\Gamma(x), x > 0$ the former expresión can be simplified as follows. Consider the first ratio involved in the $rl^{th}$ term in the product above:

$$\frac{\Gamma(2n_r^{\sigma}+a+b)}{\Gamma(2(n_r+n_r^{\sigma})+a+b)}$$

let $k_{1r} \coloneqq 2(n_r+n_r^{\sigma})+a+b$, since $n_r \in \mathbb{Z}_+$ and $2n_r^{\sigma}+a+b > 0\ \forall\, r = 1,2,\ldots,R$, we can write

$$\begin{aligned} &\Gamma(2(n_r+n_r^{\sigma})+a+b) = (k_{1r}-1)\Gamma(k_{1r}-1) \\ &= (k_{1r}-1)(k_{1r}-2)\Gamma(k_{1r}-2) \\ &= (k_{1r}-1)(k_{1r}-2)\cdots(k_{1r}-2n_r)\Gamma(k_{1r}-2n_r) \\ &= \left[\prod_{i=1}^{2n_r}(2(n_r+n_r^{\sigma})+a+b-i)\right]\Gamma(2n_r^{\sigma}+a+b) \end{aligned}$$

thus,

$$\frac{\Gamma(2n_r^{\sigma}+a+b)}{\Gamma(2(n_r+n_r^{\sigma})+a+b)}=\frac{1}{\prod_{i=1}^{2n_r}(2(n_r+n_r^{\sigma})+a+b-i)}$$

proceeding similarly, it can be shown that $\forall\ r=1,2,\dots,R;\ \forall\ l=1,2,\dots,L$

$$\frac{\Gamma(n_{Alr}+n_{Alr}^{\sigma}+a)}{\Gamma(n_{Alr}^{\sigma}+a)}=\prod_{i=1}^{n_{Alr}}(n_{Alr}+n_{Alr}^{\sigma}+a-i)$$
$$\frac{\Gamma(n_{Blr}+n_{Blr}^{\sigma}+b)}{\Gamma(n_{Blr}^{\sigma}+b)}=\prod_{i=1}^{n_{Blr}}(n_{Blr}+n_{Blr}^{\sigma}+b-i)$$

consequently,

$$m_{\pi}(\boldsymbol{x}|\boldsymbol{x}_{obs})=\prod_{r=1}^{R}\prod_{l=1}^{L}\frac{\prod_{i=1}^{n_{Alr}}(n_{Alr}+n_{Alr}^{\sigma}+a-i)\prod_{i=1}^{n_{Blr}}(n_{Blr}+n_{Blr}^{\sigma}+b-i)}{\prod_{i=1}^{2n_r}(2(n_r+n_r^{\sigma})+a+b-i)}$$

**Proposition 3**. 1) The posterior predictive density is increasing in $b$ for fixed $a$, iff both are contained in the subset of its domain that together with the observed population structure satisfy the following condition

$$\sum_{(l,r)\in P_b}S_{lrb}>-\sum_{(l,r)\in N_b}S_{lrb}$$

where, $S_{lrb}=\psi(c_{1lr})-\psi(c_{2r})+\psi(c_{3r})-\psi(c_{4lr})$, $c_{1lr}:=n_{Blr}+n_{Blr}^{\sigma}+b$, $c_{2r}:=2(n_r+n_r^{\sigma})+a+b$, $c_{3r}:=2n_r^{\sigma}+a+b$, $c_{4lr}:=n_{Blr}^{\sigma}+b$, $N_b=\{(l,r)\colon S_{lrb}<0\}$, $P_b=\{(l,r)\colon S_{lrb}>0\}$ and $\psi(\cdot)$ is the digamma function (Artin, 1964). Also, $n_{Alr}>0$ is a necessary condition for $S_{lrb}<0$, that is, allele $B$ of the $l^{th}$ marker cannot be fixed in the sample from the $r^{th}$ subpopulation. Moreover, the posterior predictive density is decreasing in $b$ for fixed $a$ in the subset of its domain that along with the population structure meet the following necessary and sufficient condition: $\sum_{(l,r)\in P_b}S_{lrb}<-\sum_{(l,r)\in N_b}S_{lrb}$.
2) The posterior predictive density is increasing in $a$ for fixed $b$, iff both are contained in the subset of its domain that along with the observed population structure satisfy

$$\sum_{(l,r)\in P_a}S_{lra}>-\sum_{(l,r)\in N_a}S_{lra}$$

where, $S_{lra}=\psi(d_{1lr})-\psi(c_{2r})+\psi(c_{3r})-\psi(d_{4lr})$, $d_{1lr}:=n_{Alr}+n_{Alr}^{\sigma}+a$, $d_{4lr}:=n_{Alr}^{\sigma}+a$, $N_a=\{(l,r)\colon S_{lra}<0\}$, $P_a=\{(l,r)\colon S_{lra}>0\}$ with $n_{Blr}>0$ being a necessary condition for $S_{lra}<0$, that is, allele $A$ of the $l^{th}$ marker cannot be fixed in the sample from the $r^{th}$ subpopulation. Moreover, the posterior predictive density is decreasing in $a$ for fixed $b$ iff they are in the subset of the function's domain that together with the population structure satisfy $\sum_{(l,r)\in P_a} S_{lra} < -\sum_{(l,r)\in N_a} S_{lra}$.

3) Let

$$\mathcal{S}_1 := L\sum_{r=1}^{R}\sum_{i=1}^{2n_r}\frac{1}{(2(n_r+n_r^{\sigma})+a+b-i)^2}$$
$$\mathcal{S}_2 := \sum_{l=1}^{L}\sum_{r=1}^{R}\sum_{i=1}^{n_{Alr}}\frac{1}{(n_{Alr}+n_{Alr}^{\sigma}+a-i)^2}$$
$$\mathcal{S}_3 := \sum_{l=1}^{L}\sum_{r=1}^{R}\sum_{i=1}^{n_{Blr}}\frac{1}{(n_{Blr}+n_{Blr}^{\sigma}+b-i)^2}$$

The posterior predictive density function is strictly concave in the proper subset of its domain that along with the population genetic structure satisfy $\frac{1}{\mathcal{S}_3}+\frac{1}{\mathcal{S}_2}<\frac{1}{\mathcal{S}_1}$; a sufficient condition to fulfil this inequality is:

$$\mathcal{S}_1 > \frac{\mathcal{S}_2+\mathcal{S}_3}{4}$$

**Proof.**
***Part 1.***
Let $l_{m_\pi} = \ln\big(m_\pi(\boldsymbol{x}|\boldsymbol{x}_{obs})\big)$; therefore,

$$\begin{aligned}
l_{m_\pi} &= \sum_{r=1}^{R}\sum_{l=1}^{L}\Big[\ln\big(\Gamma(2n_r^{\sigma}+a+b)\big)-\Big(\ln\big(\Gamma(n_{Alr}^{\sigma}+a)\big)+\ln\big(\Gamma(n_{Blr}^{\sigma}+b)\big)\Big) \\
&\quad + \ln\big(\Gamma(n_{Alr}+n_{Alr}^{\sigma}+a)\big)+\ln\big(\Gamma(n_{Blr}+n_{Blr}^{\sigma}+b)\big) \\
&\quad - \ln\big(\Gamma(2(n_r+n_r^{\sigma})+a+b)\big)\Big] \\
&= \sum_{r=1}^{R}\sum_{l=1}^{L}\Big[\ln\big(\Gamma(f_{1lr}(a,b))\big)-\Big(\ln\big(\Gamma(g_{1lr}(a))\big)+\ln\big(\Gamma(\delta_{1lr}(b))\big)\Big)+\ln\big(\Gamma(g_{2lr}(a))\big) \\
&\quad + \ln\big(\Gamma(\delta_{2lr}(b))\big)-\ln\big(\Gamma(f_{2lr}(a,b))\big)\Big]
\end{aligned}$$

where

$$\begin{aligned}
f_{1lr}(a,b) &= 2n_r^{\sigma}+a+b \\
f_{2lr}(a,b) &= 2(n_r+n_r^{\sigma})+a+b \\
g_{1lr}(a) &= n_{Alr}^{\sigma}+a \\
g_{2lr}(a) &= n_{Alr}+n_{Alr}^{\sigma}+a
\end{aligned}$$

$$\delta_{1lr}(b) = n_{Blr}^{\sigma} + b$$
$$\delta_{2lr}(b) = n_{Blr} + n_{Blr}^{\sigma} + b$$

Using the chain rule, notice that for $j = 1,2$

$$\frac{\partial}{\partial a}\ln\left(\Gamma\left(f_{jlr}(a,b)\right)\right) = \psi\left(f_{jlr}(a,b)\right)\frac{\partial}{\partial a}\left(f_{jlr}(a,b)\right)$$
$$= \psi\left(f_{jlr}(a,b)\right)$$
$$\frac{\partial}{\partial b}\ln\left(\Gamma\left(f_{jlr}(a,b)\right)\right) = \psi\left(f_{jlr}(a,b)\right)\frac{\partial}{\partial b}\left(f_{jlr}(a,b)\right)$$
$$= \psi\left(f_{jlr}(a,b)\right)$$
$$\frac{d}{da}\ln\left(\Gamma\left(g_{jlr}(a)\right)\right) = \psi\left(g_{jlr}(a)\right)\frac{d}{da}\left(g_{jlr}(a)\right)$$
$$= \psi\left(g_{jlr}(a)\right)$$
$$\frac{d}{db}\ln\left(\Gamma\left(\delta_{jlr}(b)\right)\right) = \psi\left(\delta_{jlr}(b)\right)\frac{d}{db}\left(\delta_{jlr}(b)\right)$$
$$= \psi\left(\delta_{jlr}(b)\right)$$

where $\psi(\cdot)$ is the digamma function (Artin, 1954). Consequently, the $lr^{th}$ summand of $\frac{\partial}{\partial a} l_{m_\pi}$ is of the form:

$$S_{lra} = \psi(2n_r^{\sigma} + a + b) - \psi(2(n_r + n_r^{\sigma}) + a + b) + \psi(n_{Alr} + n_{Alr}^{\sigma} + a) - \psi(n_{Alr}^{\sigma} + a)$$

Using hte following recurrence property

$$\psi(x+1) = \frac{1}{x} + \psi(x+1), \qquad x > 0$$

$S_{lra}$ can be expressed in a "digamma free" form. To this end, let $k_{1lr} \coloneqq n_{Alr} + n_{Alr}^{\sigma} + a$; since $n_{Alr}, n_{Alr}^{\sigma} \in \mathbb{Z}_+,\ a > 0$

$$\psi(k_{1lr}) = \frac{1}{k_{1lr} - 1} + \psi(k_{1lr} - 1)$$
$$= \frac{1}{k_{1lr} - 1} + \frac{1}{k_{1lr} - 2} + \psi(k_{1lr} - 2)$$
$$= \sum_{i=1}^{n_{Alr}} \left(\frac{1}{k_{1lr} - i}\right) + \psi(k_{1lr} - n_{Alr})$$
$$= \sum_{i=1}^{n_{Alr}} \left(\frac{1}{n_{Alr} + n_{Alr}^{\sigma} + a - i}\right) + \psi(n_{Alr}^{\sigma} + a)$$

similarly,

$$\psi(2(n_r+n_r^{\sigma})+a+b)=\sum_{i=1}^{2n_r}\left(\frac{1}{2(n_r+n_r^{\sigma})+a+b-i}\right)+\psi(2n_r^{\sigma}+a+b)$$

and consequently

$$S_{lra}=\sum_{i=1}^{n_{Alr}}\left(\frac{1}{n_{Alr}+n_{Alr}^{\sigma}+a-i}\right)-\sum_{i=1}^{2n_r}\left(\frac{1}{2(n_r+n_r^{\sigma})+a+b-i}\right)$$

Analogously, it can be found that

$$\frac{\partial}{\partial b}l_{m_\pi}=\sum_{l=1}^{L}\sum_{r=1}^{R}S_{lrb}$$
$$S_{lrb}\coloneqq\sum_{i=1}^{n_{Blr}}\left(\frac{1}{n_{Blr}+n_{Blr}^{\sigma}+b-i}\right)-\sum_{i=1}^{2n_r}\left(\frac{1}{2(n_r+n_r^{\sigma})+a+b-i}\right)$$

Notice that studying each summand in $\frac{\partial}{\partial a}l_{m_\pi}$ and $\frac{\partial}{\partial b}l_{m_\pi}$ is usfeul to analyze the behavior of $m_\pi(\boldsymbol{x}|\boldsymbol{x}_{obs})$ as a function of $a$ and $b$. Then, let us focus on $S_{lrb}$ and let

$$\begin{aligned}
c_{1lr} &\coloneqq n_{Blr}+n_{Blr}^{\sigma}+b\\
c_{2r} &\coloneqq 2(n_r+n_r^{\sigma})+a+b\\
c_{3r} &\coloneqq 2n_r^{\sigma}+a+b\\
c_{4lr} &\coloneqq n_{Blr}^{\sigma}+b
\end{aligned}$$

hence

$$S_{lrb}=\psi(c_{1lr})-\psi(c_{2r})+\psi(c_{3r})-\psi(c_{4lr})$$

**Remark 1.** Consider the following facts: $\psi(\cdot)$ is a monotonically increasing concave function (Artin, 1954) and $c_{2r}>c_{1lr}$, $c_{3r}>c_{4lr}$, $c_{1lr}\geq c_{4lr}$ and $c_{2r}>c_{3r}$ $\forall\, l=1,2,\dots,L,\ \forall\, r=1,2,\dots,R$. This leads to

$$\begin{aligned}
\psi(c_{1lr})-\psi(c_{2r})<0\ \forall\, a,b>0\\
\psi(c_{3r})-\psi(c_{4lr})>0\ \forall\, a,b>0
\end{aligned}$$

Therefore, the sign of $S_{lrb}$ depends on the structure of the observed and "future" subpopulation $r$ for locus $l$, $a$ and $b$. There is no clear relationship between $c_{1lr}$ and $c_{3r}$; thus, there are three posible cases: $c_{1lr}<c_{3r}$, $c_{1lr}=c_{3r}$ and $c_{1lr}>c_{3r}$. The relationships are met if and only if, $n_{Blr}<n_{Alr}^{\sigma}+a$, $n_{Blr}=n_{Alr}^{\sigma}+a$ and $n_{Blr}>n_{Alr}^{\sigma}+a$, respectively; notice that the three scenarios are posible.
Let

$$\delta_1 \coloneqq c_{3r} - c_{4lr}$$
$$\delta_2 \coloneqq c_{2r} - c_{1lr}$$
$$\Delta_{1lr} \coloneqq \psi(c_{3r}) - \psi(c_{4lr})$$
$$\Delta_{2lr} \coloneqq \psi(c_{2r}) - \psi(c_{1lr})$$

therefore, $S_{lrb} = \Delta_{1lr} - \Delta_{2lr}$. Now, due to the facts pointed out in Remark 1, it follows that $\delta_1 = \delta_2$ or $\delta_1 > \delta_2$ are sufficient conditions for $\Delta_{1lr} > \Delta_{2lr}$. However, notice that

$$\delta_1 > \delta_2 \Leftrightarrow c_{3r} - c_{4lr} > c_{2r} - c_{1lr}$$
$$\Leftrightarrow 0 > 2n_r - n_{Blr}$$
$$\Leftrightarrow n_{Blr} > 2n_r \quad \rightarrow\leftarrow$$

The contradiction follows because $2n_r = n_{Alr} + n_{Blr}$; $n_{Alr}, n_{Blr} \in \mathbb{Z}_+$. As a consequence, condition $\delta_1 > \delta_2$ is not posible. Similarly,

$$\delta_1 = \delta_2 \Leftrightarrow n_{Blr} = 2n_r$$

which means that allele $B$ of the $l^{th}$ marker is fixed in the sample from the $r^{th}$ subpopulation. Typically, this setting is not observed because genomic data is edited in such a way that loci with fixed alleles are excluded from the análisis. Ont he other hand, $\delta_1 < \delta_2$, the most common situation, is necessary, but no sufficient for $\Delta_{1lr} < \Delta_{2lr}$, thus, if the genetic structure of the observed and "future" $r^{th}$ subpopulation for the $l^{th}$ marker and the hyperparameters are such that for the corresponding values of $c_{1lr}$, $c_{2r}$, $c_{3r}$ and $c_{4lr}$, the magnitude of

$$\delta_1 - \delta_2 = 2n_r - n_{Blr} = n_{Alr}$$

is large enough to meet $\Delta_{1lr} < \Delta_{2lr}$ then $S_{lrb} < 0$. T osee why, consider shifting $c_{1lr}$, $c_{2r}$, $c_{3r}$ and $c_{4lr}$, then the difference $\delta_1 - \delta_2 = n_{Alr}$ remains the same, but the values of $\Delta_{1lr}$ and $\Delta_{2lr}$ will change as well as their difference.
Now, let

$$N_b = \{(l,r): S_{lrb} < 0\}$$
$$Z_b = \{(l,r): S_{lrb} = 0\}$$
$$P_b = \{(l,r): S_{lrb} > 0\}$$

because $N_b$, $Z_b$ and $P_b$ are mutually disjoint, $\frac{\partial}{\partial b} l_{m_\pi}$ can be written as:

$$\frac{\partial}{\partial b} l_{m_\pi} = \sum_{(l,r) \in N_b} S_{lrb} + \sum_{(l,r) \in P_b} S_{lrb}$$

and consequently,

$$l_{m_\pi} \; is \; incresing \; in \; b \; \Leftrightarrow \sum_{(l,r)\,\in\,P_b} S_{lrb} > - \sum_{(l,r)\,\in\,N_b} S_{lrb}$$

***Part 2.***

Proceeding similarly top art 1, it can be shown that $n_{Alr} > 0$ is necessary for $S_{lra} < 0$ and

$$l_{m_\pi} \; is \; incresing \; in \; b \; \Leftrightarrow \sum_{(l,r)\,\in\,P_a} S_{lra} > - \sum_{(l,r)\,\in\,N_a} S_{lra}$$

where

$$N_a = \{(l,r): S_{lra} < 0\}$$
$$P_a = \{(l,r): S_{lra} > 0\}$$

***Part 3.***

Now let´s focus on the Hessian matrix of the log-posterior predictive distribution seen as a function of $(a, b)$.

$$H_{l_m} = \begin{bmatrix} \frac{\partial^2}{\partial a^2} l_{m_\pi} & \frac{\partial^2}{\partial a \partial b} l_{m_\pi} \\ \frac{\partial^2}{\partial b \partial a} l_{m_\pi} & \frac{\partial^2}{\partial b^2} l_{m_\pi} \end{bmatrix}$$

by Clairut´s theorem, $\frac{\partial^2}{\partial b \partial a} l_{m_\pi} = \frac{\partial^2}{\partial a \partial b} l_{m_\pi}$, thus

$$|H_{l_m}| = \left(\frac{\partial^2}{\partial a^2} l_{m_\pi}\right)\left(\frac{\partial^2}{\partial b^2} l_{m_\pi}\right) - \left(\frac{\partial^2}{\partial a \partial b} l_{m_\pi}\right)^2$$

now, using the “digamma free” form of the first derivatives it follows that

$$\frac{\partial^2}{\partial a^2} l_{m_\pi} = \sum_{l=1}^{L}\sum_{r=1}^{R}\left(\sum_{i=1}^{2n_r} \frac{1}{(2(n_r + n_r^\sigma) + a + b - i)^2} - \sum_{i=1}^{n_{Alr}} \frac{1}{(n_{Alr} + n_{Alr}^\sigma + a - i)^2}\right)$$
$$\frac{\partial^2}{\partial b^2} l_{m_\pi} = \sum_{l=1}^{L}\sum_{r=1}^{R}\left(\sum_{i=1}^{2n_r} \frac{1}{(2(n_r + n_r^\sigma) + a + b - i)^2} - \sum_{i=1}^{n_{Blr}} \frac{1}{(n_{Blr} + n_{Blr}^\sigma + b - i)^2}\right)$$
$$\frac{\partial^2}{\partial a \partial b} l_{m_\pi} = \sum_{l=1}^{L}\sum_{r=1}^{R}\sum_{i=1}^{2n_r} \frac{1}{(2(n_r + n_r^\sigma) + a + b - i)^2}$$
$$= L \sum_{r=1}^{R}\sum_{i=1}^{2n_r} \frac{1}{(2(n_r + n_r^\sigma) + a + b - i)^2}$$

To ease notation let

$$S_{1r} = \sum_{i=1}^{2n_r} \frac{1}{(2(n_r + n_r^{\sigma}) + a + b - i)^2}$$
$$S_{2lr} = \sum_{i=1}^{n_{Alr}} \frac{1}{(n_{Alr} + n_{Alr}^{\sigma} + a - i)^2}$$
$$S_{3lr} = \sum_{i=1}^{n_{Blr}} \frac{1}{(n_{Blr} + n_{Blr}^{\sigma} + b - i)^2}$$

hence

$$\left(\frac{\partial^2}{\partial a^2} l_{m_\pi}\right)\left(\frac{\partial^2}{\partial b^2} l_{m_\pi}\right) = \left(\sum_{l=1}^{L}\sum_{r=1}^{R} S_{1r} - S_{2lr}\right)\left(\sum_{l=1}^{L}\sum_{r=1}^{R} S_{1r} - S_{3lr}\right)$$
$$= \left(L\sum_{r=1}^{R} S_{1r}\right)^2 - \left(L\sum_{r=1}^{R} S_{1r}\right)\left(\sum_{l=1}^{L}\sum_{r=1}^{R} S_{2lr} + S_{3lr}\right) + \left(\sum_{l=1}^{L}\sum_{r=1}^{R} S_{2lr}\right)\left(\sum_{l=1}^{L}\sum_{r=1}^{R} S_{3lr}\right)$$

and

$$\left(\frac{\partial^2}{\partial a \partial b} l_{m_\pi}\right)^2 = \left(L\sum_{r=1}^{R} S_{1r}\right)^2$$

thus,

$$\left|H_{l_m}\right| = \left(\sum_{l=1}^{L}\sum_{r=1}^{R} S_{2lr}\right)\left(\sum_{l=1}^{L}\sum_{r=1}^{R} S_{3lr}\right) - \left(L\sum_{r=1}^{R} S_{1r}\right)\left(\sum_{l=1}^{L}\sum_{r=1}^{R} S_{2lr} + S_{3lr}\right)$$

To further simplify notation, let

$$\mathcal{S}_1 \coloneqq L\sum_{r=1}^{R} S_{1r}$$
$$\mathcal{S}_2 \coloneqq \sum_{l=1}^{L}\sum_{r=1}^{R} S_{2lr}$$
$$\mathcal{S}_3 \coloneqq \sum_{l=1}^{L}\sum_{r=1}^{R} S_{3lr}$$

therefore,

$$\left|H_{l_m}\right| > (<) 0 \Leftrightarrow \mathcal{S}_2\mathcal{S}_3 - \mathcal{S}_1(\mathcal{S}_2 + \mathcal{S}_3) > (<) 0$$

$$\Leftrightarrow \frac{\mathcal{S}_2\mathcal{S}_3}{\mathcal{S}_2+\mathcal{S}_3} > (<)\mathcal{S}_1$$
$$\Leftrightarrow \frac{1}{\mathcal{S}_2}+\frac{1}{\mathcal{S}_3} > (<)\mathcal{S}_1$$

Following the same rationale presented in the proof of Proposition 2 (Additional file 1), because $\dim(H_{l_m}) = 2 \times 2$, if $|H_{l_m}| < 0$ then $H_{l_m}$ is indefinite. Furthermore, becasue $H_{l_m}$ is symmetric, a necessary condition for its determinant to be positive is that its diagonal elements $\mathcal{S}_1 - \mathcal{S}_2$ and $\mathcal{S}_1 - \mathcal{S}_3$ have the same sign. Thus, if $|H_{l_m}| > 0$, then either $\mathcal{S}_1 - \mathcal{S}_2 > 0$ and $\mathcal{S}_1 - \mathcal{S}_3 > 0$ or $\mathcal{S}_1 - \mathcal{S}_2 < 0$ and $\mathcal{S}_1 - \mathcal{S}_3 < 0$, but the first case implies that:

$$\frac{1}{\mathcal{S}_1} < \frac{1}{\mathcal{S}_2} \ and \ \frac{1}{\mathcal{S}_1} < \frac{1}{\mathcal{S}_3}$$
$$\Leftrightarrow \frac{1}{\mathcal{S}_2}+\frac{1}{\mathcal{S}_3} > \frac{2}{\mathcal{S}_1} > \frac{1}{\mathcal{S}_1}$$

Which,from the results above implies that $|H_{l_m}| < 0$ leading to a contradiction. Now, if $|H_{l_m}| > 0$ it was already proven that it amounts to $\frac{1}{\mathcal{S}_2}+\frac{1}{\mathcal{S}_3} > \frac{1}{\mathcal{S}_1}$, so

$$\frac{1}{\mathcal{S}_2}+\frac{1}{\mathcal{S}_3} < \frac{2}{\mathcal{S}_1}$$

which is congruent with the second case presented above, i.e., $\mathcal{S}_1 - \mathcal{S}_2 < 0$ and $\mathcal{S}_1 - \mathcal{S}_3 < 0$, that is, both diagonal entries of $H_{l_m}$ are negative. In summary, if $|H_{l_m}| > 0$ the the diagonal entries of $H_{l_m}$ necessarily have to be negative, so in light of Theorem 14.9.5 in Harville (1997) it follows that $H_{l_m}$ cannot be positive and its first order principal minor is negative, consequently it is negative definite. Therefore, the posterior predictive density is logarithmically concave and, hence, concave in the set of $a, b > 0$ that along with the population structure satisfy the condition $\frac{1}{\mathcal{S}_2}+\frac{1}{\mathcal{S}_3} < \frac{1}{\mathcal{S}_1}$.

Finally, by corollary 3.2, $\mathcal{S}_1 < \mathcal{S}_2 + \mathcal{S}_3$, which implies

$$\frac{1}{\mathcal{S}_2+\mathcal{S}_3} < \frac{1}{\mathcal{S}_1}$$

but, corollary 3.1 implies that,

$$\frac{1}{\mathcal{S}_2+\mathcal{S}_3} < \frac{1}{\mathcal{S}_2}+\frac{1}{\mathcal{S}_3}$$

this shows that $\frac{1}{\mathcal{S}_2}+\frac{1}{\mathcal{S}_3} < \frac{1}{\mathcal{S}_1}$ defines a proper subset of the function domain.

Now, if $\mathcal{S}_1 > \frac{\mathcal{S}_2+\mathcal{S}_3}{4}$, then

$$\frac{1}{\mathcal{S}_1} < \frac{4}{\mathcal{S}_2 + \mathcal{S}_3}$$

which, upon application of corollary 3.1 yields

$$\frac{1}{\mathcal{S}_1} < \frac{4}{\mathcal{S}_2 + \mathcal{S}_3} < \frac{1}{\mathcal{S}_2} + \frac{1}{\mathcal{S}_3}$$

∎